\documentclass{aa}
\usepackage{txfonts}
\usepackage{graphics}
\usepackage{latexsym}
\newcommand{\teff}{$T_{\rm eff} \ $}

\newcommand{\dex}{\,dex}

\begin{document} 
\title{A grid of MARCS model atmospheres for late-type stars I.\\ 
       Methods and general properties}

\titlerunning{A grid of MARCS model atmospheres}

\author{
        Bengt Gustafsson\inst{1}
   \and Bengt Edvardsson\inst{1}
   \and Kjell Eriksson\inst{1}
   \and Uffe Gr\aa e J\o rgensen\inst{2}
   \and \AA ke Nordlund\inst{2}
   \and Bertrand Plez\inst{3,1}
       }

\authorrunning{Gustafsson et al.}

\institute{Department of Physics and Astronomy, Uppsala Astronomical
           Observatory, Box 515, S-751\,20 Uppsala, Sweden
     \and  Niels Bohr Institute for Astronomy, Physics and Geophysics,
           Copenhagen University, Blegdamsveg 17, Copenhagen {\rm \O}, DK-2100, Denmark 
     \and  GRAAL, Universit\'e de Montpellier II, F-34095 Montpellier Cedex 05, France}

\date{Received <date> / Accepted <date>}
\offprints{Bengt Gustafsson,
\email{Bengt.Gustafsson@astro.uu.se}}

\abstract{ 
In analyses of stellar spectra and colours, and for the analysis of integrated 
light from galaxies, a homogeneous grid of model atmospheres of late-type stars and corresponding
flux spectra is needed.}
{To construct an extensive grid of spherically symmetric models (supplemented with plane-parallel ones for the 
highest surface gravities) , built on up-to-date atomic and molecular
data, and make it available for public use.}
{The most recent version of the MARCS program is used.}
{We present a grid of about $10^4$ model atmospheres for stars with 
$2500\,{\rm K} \le\,$\teff$ \le 8000\,{\rm K}$, $-1 \le\,$log$g\,=\,$log$(GM/R^2) \le 5$\,(cgs) 
with various masses and radii, $-5 \le$[Me/H]$\le +1$, with 
$[\alpha/$Fe] = 0.0 and 0.4
and different choices of C and N abundances, including "CN-cycled" models with 
C/N = 4.07 (solar), 1.5 and 0.5, C/O ranging from 0.09 to (normally) 5.0 to also represent stars of spectral types 
R, S and N, and with $1.0 \le \xi_{\rm t} \le 5$\,km/s. 
We also list thermodynamic quantities ($T$, $P_{\rm g}$, $P_{\rm e}$, $\rho$, partial pressures 
of molecules, etc) and provide them on the World Wide Web, 
as well as calculated fluxes in approximately 108,000 wavelength points. 
Underlying assumptions in addition to 1D stratification (spherical or 
plane-parallel) include hydrostatic equilibrium, mixing-length convection and LTE.
A number of general properties of the models are discussed,
in particular in relation to the effects of changing abundances, of blanketing and of sphericity. 
We have found and illustrate positive and negative feed-backs between sphericity and molecular blanketing. 
Models are compared with other available grids and excellent
agreement is found with plane-parallel models of Castelli and Kurucz 
(if convection is treated consistently) within the 
overlapping parameter range. Although there
are considerable departures from the spherically symmetric NextGen models, the agreement with more recent
PHOENIX models is gratifying.}
{The models of the grid show considerable regularities, but some interesting departures from 
general patterns occur for the coolest models due to the molecular opacities. We have tested a number of approximate "rules of thumb" 
concerning effects of blanketing and sphericity and found them to often be astonishingly accurate. 
Some interesting new phenomena have been discovered and explored, such as the intricate coupling between blanketing
and sphericity, and the strong effects of carbon enhancement on metal-poor models. 
We give further details of line absorption data for molecules, as well as details of models and comparisons with 
observations in subsequent papers.}


\maketitle 

%%%%%%%%%%%%%%%%%%%%%%%%%%%%%%%%%%%%%%%%%%%%%%%%%%%%%%%%%%%%%%%%%%%%%%%%%%%%%%%

\section{Introduction}

Since the first grids of line-blanketed model-atmospheres for late-type 
stars were published (Carbon \& Gingerich 1969; Querci, Querci \& Tsuji 1974; 
Peytremann 1974; Gustafsson et al. 1975; Kurucz 1979; 
Johnson, Bernat \& Krupp 1980), there has been a very impressive 
improvement in underlying data, in particular for atomic and molecular 
absorption. Accurate continuous absorption coefficients for a number 
of heavy elements have been calculated  within the Opacity Project 
(Seaton et al. 1994), and the Iron Project (Bautista 1997).
The Opacity Project also contributed accurate transition probabilities 
for a wealth of spectral lines for elements where the LS coupling 
approximation was applicable. Thanks to systematic efforts by R. Kurucz 
(see {\it http://kurucz.harvard.edu}) line lists with transition probabilities for 
millions of lines of heavy elements have been calculated. Also, the 
admirable and systematic work by experimental physicists (e.g. Blackwell et al. 1989, Nave et al. 1994, 
Hartman et al. 2003, Lawler et al. 2007) has led to the identification and measurements 
of many more metal transitions, e.g. from Fe\,{\sc i} and Fe\,{\sc ii}, 
than existed before. 
The results of these efforts have been made easily accessible in data bases, 
such as the VALD (see Piskunov et al. 1995, Stempels, Piskunov \& Barklem 2001, 
see also http://www.astro.uu.se/$\sim$vald). Additional significant progress has been made in the calculation 
of accurate damping constants for strong atomic lines (Anstee \& O'Mara 1995, Barklem, Piskunov
\& O'Mara 2000a) as well as for hydrogen self broadening (Barklem, Piskunov \&
O'Mara 2000b). 

Also for molecular absorption, impressive progress has 
been made, starting from practically nothing 30 years ago (see J\o rgensen 1994b and J\o rgensen 2005 for  
reviews). Very extensive line lists have thus been calculated for most diatomic 
and polyatomic molecules that contribute opacity in stellar atmospheres. 
The lists are based on laboratory measurements of wavelengths and $gf$ values 
with simple theoretical extensions, or are the result of more extensive 
quantum-mechanical {\it ab-initio} calculations. Even if the absorption 
cross-sections are not always of satisfactory quality as yet, the existence 
of these lists makes it possible to calculate models for, e.g., M and C stars 
that are realistic enough to compare reasonably well with observations. 

Simultaneously with these improvements of basic physics, steps to improve 
the physical consistency of the models have also been taken. The traditional 
assumptions of plane-parallel stratification in homogenous layers, of 
stationary hydrostatic equlibrium, mixing-length convection and LTE have 
stepwise been possible to relax. Grids of spherically symmetric models 
for giants and supergiants were made by Plez, Brett \& Nordlund (1992) and 
J\o rgensen, Johnson \& Nordlund (1992). Dynamic pulsating model atmosheres for 
cool giants have been developed by Wood (1979) and Bowen (1988), and these were
improved with time-dependent dust formation by Fleischer et al. (1992) and  
H\"ofner \& Dorfi (1997), and further developed to include frequency-dependent
radiative transfer by H\"ofner (1999) and H\"ofner et al. (2003). 
Nordlund \& Stein and collaborators (Nordlund 1982, 
Nordlund \& Dravins 1990, Stein \& Nordlund 1998, Asplund et al. 1999, 2000a) 
developed 3D simulations with proper hydrodynamics and radiation fields 
taken into acount for solar-type stars, and 
Freytag (2001) developed full "star-in-a-box" models for supergiants. 
These simulations show a striking agreement with observations 
of solar granulation and spectral line profiles for solar type stars, and 
clearly demonstrate the qualitative difference between traditional 1D models 
and reality: while the temperature structure in the upper layers of the 1D 
models is determined by radiative cooling and heating, the radiative heating 
in real late-type stars is balanced to a significant extent also by expansion 
cooling of uprising gas. 

A consistent treatment of radiative transfer in models for late-type stars 
without making the assumption of LTE is complicated, due to the great number of 
atomic and molecular species affecting the radiative field, the wealth of 
levels and transitions in these species, and the lack of basic data for these 
transitions -- not the least the lack of cross sections for atomic and molecular 
collisions with electrons and hydrogen atoms. From the first attempt to
construct a reasonably realistic NLTE model for a late-type star (the Sun) made
by Anderson (1989), the development of algorithms and computers has now made 
it possible to calculate grids of such models (Hauschildt et al. 2002).
The lack 
of accurate collision cross sections, however, is still a major problem in these
efforts.

In the present paper we present an extensive grid of model atmospheres for late-type
stars. The models are still classical, in the sense that they are one-dimensional, i.e. spherically
symmetric or plane-parallel, with LTE assumed and convection taken into account by using the
standard mixing-length theory. In spite of the inadequacy of these underlying physical 
assumptions, which must be abandoned in accurate work as soon as 
possible, we still think the present grid will be useful for years to come, 
for various applications extending from studies of individual stars to galactic 
evolution and populations in external galaxies.

The grid extends from models for  A-type stars to M, S and C 
star models, from dwarfs to supergiants in luminosity, and from  
$10^{-5}$ times solar to 10 times solar in metallicity, and with various choices 
of parameters like radius and mass, abundances of C, N and O, and of 
``$\alpha$ elements" (Ne, Mg, Si, S, Ar, Ca, Ti) relative to Fe, as well as of
microturbulence parameter.
We also provide model Spectral Energy Distributions (SEDs), sampled in about $10^5$ frequency 
points. As regards methods and underlying data used, the model grid is 
homogeneous.  

The grid is presented in 
a number of papers. In this initial paper we first (Sec. 2) sketch the development of our computer
code MARCS, from which models produced at different earlier stages are in wide use, 
give a general outline of the methods now used (Sec. 3 and 4), 
and underlying data (Sec. 5), as well as properties of the grid (Sec. 6). Models are also 
compared with models from other contemporary grids (Sec. 7). In subsequent papers 
we shall discuss models for A-G stars (Edvardsson et al., Paper II), K and R 
stars (Eriksson et al., Paper III), M stars (Plez et al., Paper IV), 
S stars (Plez et al., Paper V), C stars (J\o rgensen et al., Paper VI) and models for 
very metal-poor stars (Paper VII). In 
these subsequent papers, the most important new opacity data are detailed, the model 
structures are described and analysed, and the model SED:s, fluxes and colours 
are compared to some observational data in order to explore and illustrate the 
applicability of the models. 

The total number of models is about $10^4$. The details of these models, 
including their calculated fluxes at different wavelengths, 
are available via {\it http://marcs.astro.uu.se}. 

\section{The development of MARCS}

Since the early 1970ies, we have developed and used the code MARCS 
for constructing late-type model atmospheres. Spectral-line blanketing was 
first considered using Opacity Distribution Functions (Gustafsson et al. 1975) 
and this technique and its underlying so-called ODF approximation was proven 
to be fully adequate for F, G and K stars. Thus, models could be calculated 
with just a few hundred frequency points and extensive grids for G and K giants
(Bell et al. 1976) as well as R stars (Olander 1981) were issued on the basis of
the ODF approximation. The ODFs were 
constructed with a program called SSG, originally devised for synthetic colour 
calculations by R.A. Bell (see, e.g. Bell 1971). The line list used by this 
program was based on laboratory wavelengths and oscillator strenghs, as well as 
``astrophysical oscillator strenghts''  derived from the solar spectrum. 
The total list contained about 50,000 lines. The SSG program was next used 
to calculate model spectra and colours, which were systematically compared 
with observations (e.g. Gustafsson \& Bell 1979; Bell \& Gustafsson 1989).
From these comparisons a generally good agreement was found, which made us 
confident in applying the models in a number of studies, e.g. of abundances 
in globular cluster stars. 
However, there was a mismatch in certain wavelength 
regions, in particular in the ultraviolet and violet, where the models were 
obviously too bright. We tentatively ascribed this discrepancy to the absence 
in the line lists and the models of many weak lines, which together affect the 
stellar spectra.

There were also more fundamental problems with the ODF method. 
One was that the ODF:s had to be recalculated when the chemical composition
was changed, which made it inflexible for stars with special and peculiar
abundances, such as carbon stars. To circumvent this, a special method to add
ODF:s for individual molecules was invented (Saxner \& Gustafsson, 1984).
However, for N-type carbon star models we also found that the effects 
of molecules in the upper atmospheric layers, contributing strong opacity at 
wavelengths different from the atomic absorption at greater dephts, made the ODF 
approximation unsatisfactory (cf. Ekberg et al. 1986). To provide satisfactory 
analysis for a study of the chemical composition of carbon stars (Lambert et 
al. 1986), we consequently changed the scheme to Opacity Sampling, in which 
the adequate absorption at each monochromatic wavelength point is treated in 
full detail (yet, assuming LTE in our case), following Peytremann (1974) and Sneden,
Johnson \& Krupp (1976). In order to obtain an accurate 
representation of the total radiation contribution to the heat balance and the 
radiative pressure force this required an increase of the number of frequency 
points by one to two orders of magnitude. A number of model grids were calculated
with this new program, for carbon stars (cf. Lambert et al. 1986; J\o rgensen, 
Johnsson \& Nordlund 1992), for M giants (Plez, Brett \& Nordlund 1992) and M dwarfs (Brett
\& Plez, 1993) and 
applied to studies of individual stars. We gradually extended the underlying 
line list by calculating and adopting more complete data for diatomic molecules,
such as TiO (J\o rgensen 1994a, Plez, Brett \& Nordlund, 1992, Plez 1998), 
as well as polyatomic molecules like 
HCN (J\o rgensen et al, 1985, J\o rgensen 1990), C$_2$H$_2$ (J\o rgensen thesis, cited
in J\o rgensen 1989) 
and H$_2$O (Plez, Brett \& Nordlund 1992, Alvarez and \& Plez 1998, 
J\o rgensen et al. 2001, Decin et al. 2000).

In order to provide models for a major effort to study the build-up of chemical 
elements in the Galactic Disk using solar-type stars, we further extended the 
line list, by now adding the very great number of metal lines from Kurucz's 
calculations (Plez, Brett \& Nordlund 1992; Edvardsson et al. 1993).  This 
version of the code was also used in calculating a grid for hydrogen-poor carbon stars 
(R Cr B stars, Asplund et al. 1997), and applied in an abundance study of such 
stars (Asplund et al. 2000b). Somewhat different versions of the code were 
used  to compare models to ISO spectra and other IR spectra, partly for 
calibration purposes of the satellite instrumentation, of F, G and K stars 
(van der Bliek, Gustafsson \& Eriksson, 1996, and subsequent papers, 
Decin et al. 2003 and references therein) as well as of M stars (Alvarez, 
Lanon, Plez et al. 2000. Fluks et al. 1994) and carbon stars (Loidl, Lancon \& J\o rgensen 2001).
We also made comparisons of model colours with observed ones (Bessell, Castelli 
\& Plez, 1998), and applied colours and SEDs for
establishing temperature scales (see, e.g., Massey et al. 2007 and references
therein). 

The development of MARCS has always been driven by our own (and collaborators') 
needs for models for studies of particular stars, mainly for abundance 
determination. It has been a flexible ``laboratory set-up'', rather than a 
``common-user instrument'' , and we have been reluctant to calculate models for 
stars of types that we cannot explore ourselves to understand the limitations of
our models. Until now, less than $10^3$ models have been published, which 
is a small number compared to the number actually calculated and used. Also, several different
versions of MARCS have appeared, which has led to some confusion because of inhomogeneities
among the grids produced. It was therefore judged important to develop a common 
updated version, and to use it to construct and publish an extensive homogeneous grid. 
The results of this effort is presented in this and subsequent papers.

\section{Physical assumptions and equations}

The basic assumptions have been listed above. We shall not discuss their 
adequacy further here (for a review, see e.g. Gustafsson \& J\o rgensen 1994), 
but comment on them one at a time, and in connection with that list some of 
the corresponding fundamental equations, for further reference below. 
 
\subsection{Stratification and hydrostatic equilibrium}

Assuming spherical symmetry we may write the equation of hydrostatic equilbrium:     
\begin{eqnarray}
\label{eq1}
\nabla P_{\rm tot} = -\rho {{GM_{\rm r}}\over{r^2}}\,,
\end{eqnarray}
$P_{\rm tot}$ being the total pressure, $\rho$ the matter density, $G$ Newton's constant
of gravity and $M_{\rm r}$ the stellar
mass inside radius $r$. 
We neglect the atmospheric mass in comparison with the total stellar mass $M$ 
and thus assume $M_{\rm r} = M$.
For $\nabla P_{\rm tot}$ we have
\begin{eqnarray}
\label{eq2}
\nabla P_{\rm tot} = \nabla P_{\rm g} +
\nabla P_{\rm turb} + \nabla P_{\rm rad}\,,
\end{eqnarray}
where $P_{\rm g}$ and $P_{\rm turb}$ are the gas pressure and turbulent pressure, respectively,
and the last term is the force excerted by the radiation,
\begin{eqnarray}
\label{eq3}
\nabla P_{\rm rad} = -{1\over c} \int_0^\infty {(\kappa_{\lambda} +
\sigma_{\lambda})\, F_\lambda\, d\lambda}\,,
\end{eqnarray}
where $F_\lambda$ is the radiative energy flux per wavelength unit,
$\kappa_{\lambda}$ and $\sigma_{\lambda}$ are the monochromatic absorption and 
scattering coefficients, respectively, and $c$ is the speed of light. The boundary condition for
Eq.\,(\ref{eq2}) is 
\begin{eqnarray}
\label{eq4}
P_{\rm g}(r=\infty) =  P_{\rm turb}(r=\infty) = 0.0\,.
\end{eqnarray}
The formulation of the boundary condition involving the radiative force is, 
however, not quite trivial in practice; see Plez, Brett \& Nordlund (1992).
For the turbulent pressure $P_{\rm turb}$ one may write
\begin{eqnarray}
\label{eq5}
P_{\rm turb} = \beta\, \rho\, v_{\rm t}^2
\end{eqnarray}
where $\rho$ is the gas density and $v_{\rm t}$ a characteristic velocity. 
This pressure is measuring the force produced by the
kinetic movements of the gas, whether due to convective or other turbulent gas
motions. The parameter $\beta$ is $\sim 1$, with an exact value depending on
whether the motions occur more or less isotropically. With our general lack of 
knowledge about $v_{\rm t}$ it is reasonble to adopt an approximate recipe for
$P_{\rm turb}$. We begin with assuming a depth independent value of $v_{\rm t}$. 
The dominant depth variation in $P_{\rm g}={\cal R} \rho T/\mu_{\rm mol}$, ${\cal R}$ being the gas constant, 
occurs in $\rho$. 
Neglecting the variation in the temperature $T$ and in the mean molecular weight $\mu_{\rm mol}$ as well as in
$v_{\rm t}$ and in $M_{\rm r}=M$, one finds
\begin{eqnarray}
\label{eq6}
\nabla P_{\rm g} \approx -\rho\, {GM\over r^2}\,
\left({{1 - {1\over 4 \pi c}\, {\chi_{\rm F}\over G} (L/M)} \over
{1+{\beta {\mu_{\rm mol}\over {\cal R}T} v_{\rm t}^2}}}\right)\,,
\end{eqnarray}
where $\chi_{\rm F}$ is the flux-weighted mean of the extinction coefficient per gram 
and $L$ the luminosity of the star. Here, we have also assumed that the dominating fraction
of the stellar flux is carried by radiation, i.e. the approximation is most accurate for
the upper radiative zones of the models.  
One may regard the right-hand side of Eq.\,(\ref{eq6}) as the 
local effective surface gravity $g_{\rm eff}$ times $\rho$. We thus find that we can mimic the 
effects of the radiative force and/or the turbulent pressure on the models by using 
models with those effects neglected with an adjusted gravity: 
\begin{eqnarray}
\label{eq7}
g_{\rm eff} = g \,
\left({{1 - {1\over 4 \pi c}\, {\chi_{\rm F}\over G} (L/M)} \over
{1+{\beta {\mu_{\rm mol}\over {\cal R}T} v_{\rm t}^2}}} \right),
\end{eqnarray}
where
\begin{eqnarray}
\label{eq8}
g = g(r) = {GM\over r^2}.
\end{eqnarray} 
Defining the effective 
Eddington luminosity, $L_{\rm Edd}^{\rm eff}\equiv 4\pi G M c/\chi_{\rm F}$ we obtain
\begin{eqnarray}
\label{eq75}
g_{\rm eff}= g \,\left({1 - (L/L_{\rm Edd}^{\rm eff})} \over
{1+{\beta\gamma (v_{\rm t}/c_{\rm s})^2}}\right),
\end{eqnarray}
where $c_{\rm s}=\sqrt{\gamma P_g/\rho}$ is the sound spead
for an ideal gas and $\gamma$ is the adiabatic
index. Basically, we have neglected the depth variation in 
$\gamma (v_{\rm t}/c_{\rm s})^2$ in deriving this expression. 

Thus, a model with a turbulent velocity $v_{\rm t}$ may be represented by a 
model with a reduced gravity $g_{\rm eff}$ and $v_{\rm t} =0$, according to 
this recipe.
Similarly, effects 
from the radiation force may be mimicked by changes in $g$ or $M$. 
However, the 
Planck mean may vary strongly with depth and with stellar fundamental 
parameters, and one should therefore be careful in the use of Eq.\,(\ref{eq7})
for exploring the radiative effects on the atmospheric structure and, e.g., the
possible effects on mass-loss rates. For a more detailed study for red 
supergiants and AGB stars, see Gustafsson \& Plez (1992) and 
J\o rgensen \& Johnson (1992), respectively. 
We have tested the use of Eq.\,(\ref{eq7}) to simulate 
the effects of turbulent pressure for a number of models at various points in 
the grid and find that it leads to very small errors in the temperature 
structure (less than 5\,K in the temperature through all the model for 
a depth independent $v{\rm_t}$
in the interval 0 to 10 km/s.
We therefore have chosen to set $v_{\rm t} = 0$ for all grid models, and advice those 
who would have liked a different choice to use models with a different mass or 
$g$, according to the recipe given in Eq.\,(\ref{eq7}). It should be noted that
the mixing-length treatment of convection adopted here (see Sec. 2.2 below) 
leads to a rapidly varying formal convective velocity, in particular close to
the boundary where Schwarzschild stability sets in. If this variation were included
the term $\nabla P_{\rm turb} = \nabla (\beta\, \rho\, v_{\rm t}^2)$ would get
a major contribution from $\nabla v_{\rm t}$. However, more realistic simulations
of convection show that $v_{\rm t}$ varies much less with depth than $\rho$, in
accordance with what was
assumed in the derivation of Eq.\,(\ref{eq7}) above. 

The acceleration of gravity, $g$, for our spherical models is a depth-varying quantity, according
to Eq. \,(\ref{eq8}). Also the stellar energy flux $F(r)$ and \teff are varying with depth. Thus, we have
\begin{eqnarray}
\label{eq9}
F(r) = F_{\rm rad}(r) + F_{\rm conv}(r)
= \sigma_{\rm SB}\, T_{\rm eff}(r)^4\,,
\end{eqnarray}
where $F_{\rm rad}$ and $F_{\rm conv}$ are the radiative and convective flux, respectively, and
$\sigma_{\rm SB}$ is Stefan-Boltzmann's constant.
What remains constant (in stationary models) is the mass $M$ (since only a tiny fraction of
the total stellar mass resides in the atmosphere), and the luminosity, $L=4\pi r^2 F(r)$. 
We label the models by the values of \teff and $g$ at a radius $r=R_{\rm 1}$ where 
$\tau_{\rm Ross}$, the optical-depth scale based on the Rosseland mean opacity, is equal to 1.0, i.e.:
\begin{eqnarray}
\label{eq10}
T_{\rm eff} \equiv T_{\rm eff}(R_1) = \left({L\over 4 \pi R_1^2}\right)^{1\over 4},
\end{eqnarray}
\begin{eqnarray}
\label{eq11}
g \equiv {MG\over R_{\rm 1}^2}\,.
\end{eqnarray}

\subsection{Mixing length convection}

We have used the version of the Mixing-length ``theory'' as presented by 
Henyey et al. (1965).
The convective energy flux is given by
\begin{eqnarray}
\label{eq12}
F_{\rm conv} = {1\over 2}\, \rho\, C_{\rm p}\, T\, v_{\rm conv}\,
{\ell\over H_{\rm p}}\, \delta\Delta\,.
\end{eqnarray}
Here, 
\begin{eqnarray}
\label{eq13}
\delta\Delta = {\Gamma\over (1+\Gamma)}\, (\nabla_{\rm T} - \nabla_{\rm ad})\,,
\end{eqnarray}
where
\begin{eqnarray}
\label{eq14}
\Gamma = v_{\rm conv}\, \rho\, C_{\rm p}\, {1+y(\rho\, \chi_{\rm Ross}\,
\ell)^2\over 8\, \sigma_{\rm SB}\, T^3 \rho\, \chi_{\rm Ross}\, \ell}\,,
\end{eqnarray}
\begin{eqnarray}
\label{eq15}
\nabla_{\rm T} = {d {\ln} T\over d {\ln} P},
\end{eqnarray}
and $H_{\rm p}$ is the local pressure scale height,
\begin{eqnarray}
\label{eq16}
H_{\rm p} = {P\over g \rho}.
\end{eqnarray}
$\nabla_{\rm ad}$ is the adiabatic temperature gradient, $C_{\rm p}$ the specific heat at constant
pressure, $\chi_{\rm Ross}$ the Rosseland
mean opacity and $\ell$ the mixing
length. Eq.\,(\ref{eq12}) is valid if $(\nabla_{\rm T} - \nabla_{\rm ad})\ge 0.$ Also 
\begin{eqnarray}
\label{eq17}
v_{\rm conv} = {\ell\over H_{\rm p}} \sqrt{{GM\over r^2} H_{\rm p}\, Q\,
\delta\Delta/\nu}
\end{eqnarray}
with 
\begin{eqnarray}
\label{eq18}
Q = -{T\over \rho} \left({\partial \rho\over \partial T}\right)_{\rm p}\,, 
\end{eqnarray}
the derivative taken at constant thermodynamic pressure.
In addition to the mixing length parameter, $\alpha=\ell/H_{\rm p}$, there are 
two more explicit parameters in this formulation: $y$ which is related to the 
adopted temperature distribution within the convective elements and $\nu$ which
deals with the energy dissipation by the turbulent viscosity. As a standard, 
we have chosen the parameters according to the suggestions by Henyey et al.: 
$\alpha = 1.5$, $y = 0.076$ and $\nu = 8$.  It should be noted that different
choices than these are made, sometimes without being pointed out, in other 
current work. The effects of varying the convective parameters were explored 
for models of red giants by Gustafsson et al. (1975). 
However, these 
variations do not at all map the real possible range of errors due to our (inadequate) 
treatment of the convective energy transport and even less can give relevant 
information on the effects of the thermal inhomogeneities generated by 
convection. 

The convective flux is added to the radiation energy flux,
and we may then write the energy equilibrium equation
\begin{eqnarray}
\label{eq19}
F_{\rm conv} + F_{\rm rad} = {L\over 4 \pi r^2}\,.
\end{eqnarray}

\subsection{LTE and radiative transfer}

All number densities of all atoms and molecules are assumed to follow from the 
corresponding laws for thermal equilibrium,  the Saha equation and the
corresponding equation of chemical equilibrium for the molecules 
(Gibson \& Heitler 1928, see also Russell 1934). Similarly, 
all excitation equilibria and all partition functions are calculated adopting 
the Boltzmann distribution, with higher terms in the atomic partition 
functions cut according to the method of Irwin (1981).

The radiation source function is assumed to be 
\begin{eqnarray}
\label{eq20}
S_\lambda = {\kappa_{\lambda}\over \kappa_{\lambda}+\sigma_{\lambda}} B_\lambda(T)
+ {\sigma_{\lambda}\over \kappa_{\lambda}+\sigma_{\lambda}} J_\lambda,
\end{eqnarray}
where $B_\lambda(T)$ is the Planck function.
All line absorption is assumed to occur in true absorption, i.e.
\begin{eqnarray}
\label{eq21}
\kappa_{\lambda}= \kappa_{\lambda}^{\rm cont} + \kappa_{\lambda}^{\rm line}
\end{eqnarray}
where $\kappa_{\lambda}^{\rm cont}$ and $\kappa_{\lambda}^{\rm line}$ are the sums
of all continuous absorption contributions and line absorption contributions,
respectively. The mean intensity $J_\lambda$ is calculated from 
\begin{eqnarray}
\label{eq22}
J_\lambda = \int_0^1 j_\lambda (\mu) d\mu\,.
\end{eqnarray}
Here, following Feautrier (1964), we define $j_\lambda(\mu)$ as 
\begin{eqnarray}
\label{eq23}
j_\lambda(\mu) = {1\over 2} \left(I_\lambda(\mu) + I_\lambda(-\mu)\right)
\end{eqnarray}
where $I_\lambda (\mu)$ is the specific intensity in the direction specified by
$\mu = {\rm cos} \theta$, $\theta$ being the angle relative to a stellar radius.  
The equation of radiative transfer for $j_\lambda$ is 
\begin{eqnarray}
\label{eq24}
{d^2 j_\lambda\over d\tau_\lambda^2} = j_\lambda - S_\lambda
\end{eqnarray}
where $\tau_\lambda$ is measured along the ray. For a discussion of adequate 
boundary conditions for $j_\lambda$ and their implementation, see Nordlund (1984).
Knowing $j_\lambda(\mu)$, the wavelength-integrated flux in the radial direction 
can be calculated from a derivative of the second Eddington moment $K_\lambda$, 
defined by
\begin{eqnarray}
\label{eq25}
K_\lambda = \int_0^1 \mu^2 j_\lambda d\mu\,.
\end{eqnarray}
Thus, we have for the monochromatic flux 
\begin{eqnarray}
\label{eq26}
F_\lambda =  4\pi \left({\partial K_\lambda\over \partial \tau_\lambda} -
{1\over r}\, {3 K_\lambda - J_\lambda \over
\kappa_\lambda + \sigma_\lambda}\right)\,,
\end{eqnarray}
cf. Mihalas (1978, his Eq. 2.80)  and then
\begin{eqnarray}
\label{eq27}
F_{\rm rad} =  \int F_\lambda d\lambda\,,
\end{eqnarray}
which is used in the energy conservation equation, Eq.\,(\ref{eq19}).
Alternatively, the energy balance may be expressed as 
\begin{eqnarray}
\label{eq28}
{d\over dr}{\left((F_{\rm rad} + F_{\rm conv})\cdot r^2\right)} = 0\,.
\end{eqnarray}
For $F_{\rm conv}=0$ which is often the case in the upper layers of the models, this is equivalent to
\begin{eqnarray}
\label{eq29}
\int \kappa_\lambda\, \left[J_\lambda(\tau_{\rm Ross}) -
B_\lambda(T(\tau_{\rm Ross}))\right]\, d\lambda = \nonumber \\
= q_{\rm rad} - q_{\rm thermal} = 0\,,
\end{eqnarray}
with 
\begin{eqnarray}
\label{eq30}
q_{\rm rad}\equiv \int \kappa_\lambda\, J_\lambda(\tau_{\rm Ross})\, d\lambda \nonumber \\
q_{\rm thermal}\equiv \int \kappa_\lambda\, B_\lambda(T(\tau_{\rm Ross}))\, d\lambda\,. 
\end{eqnarray}
   
\section{Physical data}

The volume of physical data needed in the calculation of model atmospheres of 
late-type stars is considerable. Data are needed for the calculation of the 
ionization equilibrium of atoms and dissociation equlibrium of molecules. 
These data include chemical composition data, ionization energies 
and dissociation energies, as well as 
partition functions. Moreover, continuous absorption and scattering 
coefficients are needed.
The by far most extensive, and most demanding need, however, 
is the various data that are necessary for the proper calculation of the line 
absorption. Here, we shall briefly present the data used to calculate 
ionization-dissociation equlibria as well as continuous absorption and 
scattering.
Some more details concerning the line-absorption data
are given in Papers II - VII.

The basic chemical composition adopted is that of the Sun, as listed by
Grevesse, Asplund \& Sauval (2007). There is still some dispute on the 
C, N and O abundances adopted there (C=8.39, N=7.78 and O=8.66) and therefore
we have alternatively taken the data of Grevesse \& Sauval (1998) with 
CNO abundances higher by about 0.2 dex for the solar-metallicity sub-grid and explore the differences in
Sec. 6.3. In varying the overall metallicity of the models ([Me/H]) we have as a standard
scaled the abundances of all elements heavier than He in unison. The abundances of so-called 
$\alpha$ elements O, Ne, Mg, Si, S, Ar, Ca, and Ti tend to vary somewhat differently for
galactic stars (see, e.g. Edvardsson et al. 1993, Reddy et al. 2003, 
Ryde \& Lambert 2004, Nissen et al. 2004, Cayrel et al. 2004); a variation which, however,
is not necessarily the same for all stellar populations in the Galaxy
(see, e.g. Fuhrmann 1998 and Bensby et al. 2005) or in other galaxies 
(see Venn et al. 2004). Therefore, we offer models with two different
sets of abundances: one with a uniform scaling for each element with [Me/H] and one 
where the $\alpha$ elements have been scaled as follows: [$\alpha$/Fe]=0.4 for $-5.0\le$[Me/H]$\le -1.0$,
[$\alpha$/Fe]=$-0.4\times$[Me/H] for $-1.0 \le$[Me/H]$\le 0.0$, and 
[$\alpha$/Fe]=0.0 for [Me/H]$\ge$ 0.0. For the giant stars, the results of the First Dredge Up
of CNO processed material is known to lead to a reduced carbon abundance and a correspondingly
increased N abundance (cf. e.g., Boothroyd \& Sackmann 1999). We therefore also offer models with revised C and N such that
C/N=1.5 and 0.5 by number, respectively, as alternatives to the solar value of C/N=4.07, 
though keeping C+N constant (cf Sec 6.3). For the corresponding giant stars, the changed C/N ratio is usually
accompanied by a decrease in the isotopic ratio $^{12}$C/$^{13}C$. For the "CN processed" models we have 
therefore changed this ratio from solar to 20 and 4, respectively. Finally, for the
M stars and carbon stars we also present models with increased C abundances, such that C/O
ranges from 0.54 to 0.99 (M and S star models) to above 1.0 (models for C stars).  
   
The atomic partition functions and ionisation energies used are from
Irwin (1981) with some modifications.
The molecular equilibria were calculated using 
the partition functions and dissociation energies of Sauval (private 
communication), which is an update of Sauval \& Tatum (1984) for diatomic molecules, and
Irwin (1988, and private communication) for polyatomic molecules.
Altogether 519 molecules were included in the equilibrium calculations.
A comparison with equlibria calculated by Piskunov (private communication) 
shows a very good agreement. 

The continuous absorption from 
H\,{\sc i}, H$^-$, H$_2^-$, H$_2^+$, He\,{\sc i}, He$^-$,C\,{\sc i},
C\,{\sc ii}, C$^-$, N\,{\sc i}, N\,{\sc ii}, N$^-$, O\,{\sc i}, O\,{\sc ii},
O$^-$, Mg\,{\sc i}, Mg\,{\sc ii}, Al\,{\sc i}, Al\,{\sc ii}, Si\,{\sc i},
Si\,{\sc ii}, Ca\,{\sc i}, Ca\,{\sc ii}, Fe\,{\sc i}, and Fe\,{\sc ii}, 
as well as CH, OH, CO$^-$ and H$_2$O$^-$ were calculated from sources
according to Table\,1. Corrections were made to the data for
C\,{\sc i}, Mg\,{\sc i}, Al\,{\sc i}, Si\,{\sc i} in order to account for
the fact that TOP base photon cross sections are published with theoretical 
energy levels derived from the model atom, which depart significantly from
the much better known data from laboratory work. For these four species 
corrections were considered necessary since they significantly 
affect the UV fluxes of the models. 
We identified the individual TOP base energy levels for these species
and shifted them to the empirical energies given by NIST ({\it
http://physics.nist.gov/PhysRefData/ASD/index.html}). 
The TOP base gives the photon absorption cross section (in Mbarn) as a
function of the energy of the ejected electron.
This electron energy was converted to the energy of the absorbed photon by
adding the ionization energy from the specific electronic level.
The cross-section data for all levels was summed up assuming LTE for a
number of temperatures and plotted at the full original wavelength
resolution at different temperatures. 
Also collision-induced absorption of H\,{\sc i}+H\,{\sc i}, H\,{\sc i}+He\,{\sc i},
H$_2$+H\,{\sc i},
H$_2$+H$_2$ and H$_2$+He\,{\sc i} was included.
H$_2$-He and H$_2$-H$_2$ CIA data from Borysow et al., referred to in Table 1, are 
available from {\it http:www.stella.nbi.dk/pub/scan}. More extensive data
sets for the CIA are publicly accessible from {\it http://www.astro.ku.dk/$\sim$aborysow}.
%40 wavelength points were then selected by eye for tabulation and 
%later interpolation at the model construction.
%The limited number of wavelength points prohibits any detailed
%description of the ubiquitous transient features in the TOP base
%continuous opacity data.

Continuous electron scattering as well as Rayleigh scattering from H\,{\sc i},
H$_2$ and He\,{\sc i} was included and was assumed to be isotropic. 

\setcounter{table}{0}
\begin{table}
\label{contop}
\caption{ 
Sources of data for continuous opacities.
``b-f'' and ``f-f'' denotes bound-free and free-free processes, respectively,
and ``CIA'' is Collision Induced Absorption (or quasi-molecular absorption). "TOPbase"
refers to the TOP data base of data from the Opacity Project (Seaton et al. 1994), 
available via {\it http://vizier.u-strasbg.fr/topbase/home.html}
}
\begin{tabular}{l l}
Ion and process  & Reference \\
\hline
H\,{\sc i} b-f and f-f & Karzas \& Latter (1961)$^\ddagger$ \\
H$^-$ b-f & Wishart (1979) \\
H$^-$ f-f & Bell \& Berrington (1987) \\
He\,{\sc i} b-f & TOPbase \\
He\,{\sc i} f-f & Peach (1970) \\
He$^-$ f-f & John (1994) \\
C\,{\sc i}, C\,{\sc ii} b-f & TOPbase$^\dagger$ \\
C\,{\sc i}, C\,{\sc ii} f-f & Peach (1970) \\
C$^-$ f-f & Bell et al. (1988) \\
N\,{\sc i}, N\,{\sc ii} b-f & TOPbase \\
N$^-$ f-f & Ramsbottom et al. (1992) \\
O\,{\sc i}, O\,{\sc ii} b-f & TOPbase \\
O$^-$ f-f & John (1975a, 1975b) \\
Mg\,{\sc i}, Mg\,{\sc ii} b-f & TOPbase$^\dagger$ \\
Mg\,{\sc i} f-f & Peach (1970) \\
Al\,{\sc i}, Al\,{\sc ii} b-f & TOPbase$^\dagger$ \\
Si\,{\sc i}, Si\,{\sc ii} b-f & TOPbase$^\dagger$ \\
Si\,{\sc i} f-f & Peach (1970) \\
Ca\,{\sc i}, Ca\,{\sc ii} b-f & TOPbase \\
Fe\,{\sc i} b-f & TOPbase, Bautista (1997) \\
Fe\,{\sc ii} b-f & TOPbase \\
other metals f-f & Peach (1970), hydr. approx.\\
H$_2^+$ f-f & Mihalas (1964) \\
H$_2^-$ f-f & John (1975a, b), John \& Williams (1975) \\
CH b-f & Kurucz et al. (1987) \\
OH b-f & Kurucz et al. (1987) \\
CO$^-$ f-f & John (1975a, 1975b) \\
H$_2$O$^-$ f-f & John (1975a, 1975b) \\
H\,{\sc i}+H\,{\sc i} CIA & Doyle (1968) \\
H\,{\sc i}+He\,{\sc i} CIA & Gustafsson \& Frommhold (2001) \\
H$_2$+H\,{\sc i} CIA & Gustafsson \& Frommhold (2003) \\
H$_2$+H$_2$ CIA & Borysow et al. (2001) \\
H$_2$+He\,{\sc i} CIA & J\o rgensen et al. (2000)\\
H\,{\sc i} scattering & Dalgarno, quoted by Gingerich (1964)\\
H$_2$ scattering & Dalgarno \& Williams (1962)\\
e$^-$ scattering & Mihalas (1978)\\
He\,{\sc i} scattering & Bues \& Wehrse (1976)\\
\hline
\multicolumn{2}{l}
{$^\ddagger$modified according to the occupation probability formalism, see text} \\
\multicolumn{2}{l}
{$^\dagger$ {\it http://cdsweb.u-strasbg.fr/OP.htx}.} \\
\multicolumn{2}{l}
{The calculated wavelengths of UV and blue C\,{\sc i}, Mg\,{\sc i},}  \\
\multicolumn{2}{l}
{Al\,{\sc i} and Si\,{\sc i} absorption edges have been shifted to their} \\
\multicolumn{2}{l}
{laboratory values, see text}\\
\end{tabular}
\end{table}

Line absorption data for atoms and first ions were originally adopted from VALD-1
(Piskunov et al. 1995)  but recently data for the ten most important species were 
modified according to VALD-2 (Stempels, Piskunov \& Barklem, 2001).
The updated species are Si\,{\sc i}, Ca\,{\sc i}, Ca\,{\sc ii}, Ti\,{\sc ii},
Cr\,{\sc i}, Cr\,{\sc ii}, Mn\,{\sc ii}, Fe\,{\sc i}, Fe\,{\sc ii}, and 
Ni\,{\sc ii}. The VALD-data adopted include the very numerous line data
calculated by Kurucz \& Bell (1995, see also  {\it http://kurucz.harvard.edu}), including lines
between predicted energy levels that are not yet experimentally verified. 
The $gf$ values were, however, modified for
948 lines from VALD-1 and 1276 lines from VALD-2 in the wavelength
range 3783 to 8968\,\AA, based on fits of synthetic solar intensity spectra. These model spectra were
calculated with the Holweger \& M\"uller (1974) model and compared with the observed solar
disk-center FTS spectrum of Brault \& Neckel (1987), see Paper\,II.
The model spectrum was required to match the observed equivalent widths to within 0.1 dex in log $gf$.The 
resulting $gf$ values are listed at {\it http://marcs.astro.uu.se}.
Line absorption data for diatomic and polyatomic molecules were 
considered for species according to Table 2. Most line lists of J\o rgensen and collaborators are described
by J\o rgensen (1997) and may be obtained by anonymous ftp via {\it www.stella.nbi.dk/pub/scan}. 
Line lists for HCN, C$_2$H$_2$ and C$_3$ will be added later to this data base in connection with the paper on carbon-enriched 
stars in the present series. 
The line lists of
Plez referred to in Table 2 as "unpublished" are available via {\it http://marcs.astro.uu.se}. 
Some of these lists were especially 
tailored for the present grid; in such cases a more detailed description of 
the line list for the particular species is to be found in one of the 
subsequent papers in the present series. For the OH A-X system we did not use the data of Gillis et al. (2001)
since the Kurucz (1995b) list includes higher vibration and rotation states. 
For MgH we did not include the new data of Skory et al. (2003) for $^{24}$MgH
since we were anxious to be consistent with the MgH lines for other Mg isotopes. 

Hydrogen line and bound-free opacity, and their merging, 
were modelled using a code by Barklem ({\it http://www.astro.uu.se/$\sim$barklem/hlinop.html}), which is based on 
the occupation probability formalism of D\"appen, Anderson \& Mihalas (1987). The details of the description
of this line opacity are given by Barklem \& Piskunov (2003). 
For the atomic lines of metals, the damping 
wings were also calculated, using the 
best available data (Anstee \& O'Mara, 1995, Barklem, Piskunov \& O'Mara, 2000a and references therein, 2000b, 
Barklem \& Aspelund-Johansson 2005, and Barklem, {\it private communication}).
For weak lines where such data were missing we just adopted the Uns\"old recipe
(cf. Uns\"old, 1955, his Eq. 76.43),
with a constant enhancement factor set to 2.0 for Na\,{\sc i}, 1.3 for
Si\,{\sc i}, 1.8 for Ca\,{\sc i}, 1.4 for Fe\,{\sc i}
and 2.5 for all other species. (The damping constants given in the line list
of Kurucz, which were obtained by perturbation theory, might have been preferred to the use of
Uns{\"o}ld values.) 
For the molecules, damping was not taken into account -- the lack of proper
damping parameters makes any such attempt questionable. Also, the huge density of
molecular lines often makes the damping wings of a line less important since the
intensity of the Doppler cores of neighbouring lines dominate. 

In general, it should be noted that although the present atomic and molecular data are very extensive,
and much more complete and accurate than, e.g., a decade ago, they are far from perfect. Considerable improvements
are still needed, both by adding many more faint lines and determining line strengths more accurately. The risk 
that present models are still underblanketed must be appreciated.   

\begin{table}\caption{
\label{lineop} 
Sources of data for molecular line opacities.
}
\begin{tabular}{l l}
Species  & Reference \\
\hline
HCN vib-rot & Harris et al. (2002),\\ &J\o rgensen et al.(2001) \\
H$_2$O vib-rot & Barber et al. (2006) \\ 
C$_2$ Phillips, Swan, \\
\,\,\,\,Ballik-Ramsay & Querci et al. (1971, priv. comm.)\\
C$_3$ & J\o rgensen et al. (1989)\\
C$_2$H$_2$ & J\o rgensen (1989)\\
CH vib-rot & J\o rgensen et al. (1996)\\
CH A-X, B-X, C-X & Plez et al. (2008)\\
CN A-X, B-X & Plez (unpublished)\\
CO vib-rot & Goorvitch (1994) \\
CO A-X & Kurucz (1995) \\
CaH A-X, B-X & Plez (unpublished) \\
FeH  F$^4\Delta$-X$^4\Delta$ & Plez (unpublished) \\
MgH A-X, B$\prime$-X & Kurucz (1995a) \\
NH A-X & Kurucz (1995a) \\
OH vib-rot  & Goldman et al. (1998)  \\
OH A-X & Kurucz  (1995a) \\
SiH A-X & Kurucz  (1995a) \\
SiO vib-rot  & Langhoff \& Bauschlicher (1993) \\
TiO $\alpha, \beta, \gamma, \gamma\prime, \delta, \epsilon, \phi,$ \\
   \,\,\,\,E$^3\Pi$-B$^3\Pi$, a$^1\Delta$-f$^1\Delta$ &Plez (1998)\\
VO A-X, B-X, C-X & Plez (unpublished)\\
ZrO B$^1\Pi$-A$^1\Delta$, B$^1\Pi$-X$^1\Sigma$, \\ \,\,\,\,C$^1\Sigma$-X$^1\Sigma$
      E$^1\Phi$-A$^1\Delta$, \\ \,\,\,\,b$^3\Pi$-a$^3\Delta$, d$^3\Phi$-a$^3\Delta$,\\
      \,\,\,\,e$^3\Pi$-a$^3\Delta$, f$^3\Delta$-a$^3\Delta$,  &Plez et al. (2003) \\
\hline

\end{tabular}
\end{table}

\section{Numerical methods}
\subsection{The general method}

Equations (\ref{eq1}, \ref{eq2}, \ref{eq3}, \ref{eq12}, \ref{eq19} or
\ref{eq29}, \ref{eq26} and \ref{eq27}) with relevant boundary conditions
form a closed system of equations for determining the dependent variables
$F_{\rm conv}$, $T$, $P_{g}$, $P_{\rm rad}$, $P_{\rm turb}$, $P_{\rm e}$,
as well as quantities like $v_{\rm conv}$, $F_{\rm rad}$, or alternatively
$q_{\rm rad}$.
In addition to these quantities, we also need 
to calculate quantities characteristic (in LTE at least) of the local 
temperature and pressure: $\kappa(\lambda)$, $\sigma(\lambda)$, $H_{\rm p}$,
$Q$, $\nabla_{\rm ad}$ and $\gamma$.
The system of equations is solved on a $\tau_{\rm Ross}$ scale,
with $\kappa_{\rm Ross}$ calculated by integration over all $\sim 10^5$ wavelength 
points. This choice of depth scale leads to temperature structures that are
only moderately affected by changes in the fundamental stellar parameters. Basically, 
a standard multi-dimensional Newton-Raphson method is applied to solve the highly non-linear
system. All equations are first 
discretized, in the variables $\tau_{\rm Ross}$ and $\lambda$.
Next, all the resulting 
equations are linearized in the dependent variables listed above. The input variables
at the calculation of thermodynamic quantities and absorption coefficients are 
$T$ and $P_e$, which is a very appropriate choice as long as H$^-$ is a dominating 
opacity source. (For the coolest models, e.g. $P_H$ or $P_{\rm g}$ could have been 
more advantageous.) As a result 
of the linearization, a set of linear equations in the variations of the dependent variables 
is obtained. The coefficients of this system contain derivatives of a great 
number of quantities relative to the dependent variables. First, a starting 
model is assumed, in order to enable a first calculation of the coefficients. 
Next,  the system of linear equations is solved numerically, the resulting 
variations are applied to the dependent variables, new coefficients are 
calculated and a new solution is obtained for the variations.
The elimination scheme is basically that of Rybicki (1971), also applied
by Gustafsson \& Nissen (1972) and Gustafsson et al. (1975), with an elimination work 
that scales linearly with the number of wavelength points, $n_{\lambda}$.

\subsection{Radiative transfer}

The total radiative flux, $F_{\rm rad}$, $q_{\rm rad}$ and
$\nabla_{\rm r} P_{\rm rad}$ are non-local functionals of the 
model structure and can be calculated from a solution of the transfer 
equation, Eq.\,(\ref{eq24}).
For the model structure obtained after each iteration, we need an accurate
solution of the equation with the source function $S_{\lambda}$
given by Eq.\,(\ref{eq20}) and Eq.\,(\ref{eq22}), which then enables us to
calculate $F_{\rm rad}$ and $q_{\rm rad}$ for that structure.
These quantities are used to calculate the corrections (right hand sides) in the 
next Newton-Raphson iteration.
We also need a linearization of $F_{\rm rad}$ and
$q_{\rm rad}$ in $T$ and $P_{\rm e}$ to calculate factors, symbolically written as e.g. 
$\delta F_{\rm rad}(\tau_i)/\delta T(\tau_k)$,
or $\delta q_{\rm rad}(\tau_i)/\delta P_{\rm e}(\tau_k)$, to be used in the 
coefficient matrix in the forth-coming iteration.
Here, $\tau_i$ and $\tau_k$ are any two radial optical depths in the stellar
atmosphere. I.e., we need to estimate how the radiation 
contribution to the energy balance in each point in the atmosphere is affected 
by variations in temperature or pressure anywhere else. Both these tasks are 
accomplished with the iterative method of Nordlund (1984) which was applied to 
model atmosphere calculations by Plez, Brett \& Nordlund (1992) for M giants 
and by J\o rgensen, Johnson \& Nordlund (1992) for carbon stars. Here, we shall 
only give a brief summary for reference.

\begin{figure}
\resizebox{\hsize}{!}{\includegraphics{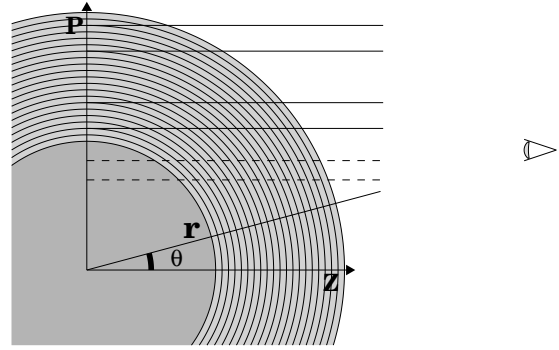}}
\caption{
Sketch of the geometric representation used for solving the spherically symmetric radiative transfer problem.
The full radius mesh used for representing the variables as functions of radius (about 7 points
per decade in $\tau_r$) is indicated along the radius vector by a set of concentric circles. The distance of rays from 
the centre of symmetry is measured by the impact parameter $p$. About 15 rays with impact parameters $\le R_c$,
the core radius, are used and are here represented by two dashed lines. The set of rays that do not hit the core, 
typically about 50 rays but dependent of wavelength, are here represented by five parallel rays. The mesh of points used for solving the
equation of radiative transfer are the crossing points between the circles and the horizontal lines in the
figure. Distances along the rays are
measured by $z=\mu r$ where $\mu=$cos$\,\theta$. 
}
\label{Sphere}
\end{figure}

The equation of radiative 
transfer is solved along a set of parallel rays, which cover a suitable range 
of impact parameters, chosen individually for each wavelength 
(cf. Fig.\,\ref{Sphere}).
In practice, 
about 15 rays equidistant in $\mu$ (the cosine of the angle relative to the 
normal) are chosen for impact parameters $< R_{\rm c}$, where $R_{\rm c}$ is
the radius for the stellar non-transparent ``core''. This core is defined by a radial 
$\tau_\lambda \sqrt{\kappa_\lambda\over \kappa_\lambda+\sigma_\lambda}> 200$, 
in order to allow a great enough thermalization depth for scattered light. 
Rays that never hit this core but
only pass through the transparent 
``atmospheric'' region, are all chosen to go through pre-selected 
$\tau_{\rm Ross}$ points in the vertical scale along a line through the stellar
centre and perpendicular to the ray. These rays are distributed with about 5\
rays per decade in $\tau_\lambda$ along this line. 

Scattering is handled by an iterative technique where the radiative transfer 
equation along a particular single ray is used to correct residual errors.
This technique uses a ``form factor'' $f_\lambda$, 
\begin{eqnarray}
\label{eq30}
f_\lambda(r) = J_\lambda/j_\lambda.
\end{eqnarray}
Here, both $J_\lambda$ and $j_\lambda$ can be taken from the previous iteration, 
since their ratio is very little dependent on the detailed shape of the source 
function; in this respect the method is analogous to the variable Eddington 
technique of Auer \& Mihalas (1970) in handling the angular dependence of the 
radiative field. After a few iterations a correct solution is obtained.
When linearizing the energy equilibrium equation (Eq.\,\ref{eq19} or \ref{eq29}) 
we linearize the transfer equation along such a representative ray and apply the 
form factor, as calculated in the preceeding iteration. This relates the 
changes in the energy balance to changes of the specific intensity along the 
ray. We adopt this relation as typical for all rays. The final result is then 
a coupling of the energy equilibrium at each point in the atmosphere to 
temperatures and electron pressures all over the model. 

The representative ray 
is chosen in the following way: At the radius in the discretization of the 
$\tau_{\rm Ross}$ scale where $\tau_\lambda$ measured radially is about 
$\sqrt{(\kappa_\lambda+\sigma_\lambda)/\kappa_\lambda}$ the set of rays hits
the sphere at a set of angles relative to the normal. We take the ray where the 
cosine for this angle is close to $1/\sqrt{3}$ -- a choice which is natural in 
view of a result of the Eddington approximation that the outgoing intensity 
from a gray atmosphere with a linear source function in this direction 
has a characteristic depth of formation which can be taken as an average depth of
formation for all disk rays.

In order to obtain a sufficient numerical accuracy we 
linearize the radiative flux directly (Eq.\,\ref{eq19}) only for optical depths 
$\tau_{\rm Ross} > 0.01$, while for smaller depths we linearize the divergence
of the flux (i.e. Eq.\,\ref{eq29}).
This avoids the loss of accuracy associated with the nearly 
constant monochromatic fluxes at most wavelengths in the optically thin layers. 

In the discretization in $\tau_{\rm Ross}$ we use 56 points,
distributed between $\tau_{\rm Ross} = 10^{-5}$ and $\tau_{\rm Ross}=100$.

The integrals over $\lambda$ are calculated using the trapezian rule. 

In the $\lambda$ discretization we use 10,000 points during the first
iterations.
In the final iterations there are 108,000 points, set to produce a sampling density
$R \equiv \lambda/\delta\lambda = 20,000$,
with $\lambda_{\rm min} = 910$\,\AA\ for models warmer than 3500\,K and
1300\,\AA\ for cooler models, and $\lambda_{\rm max} = 20\,\mu m$.
We have made a series of numerical experiments to investigate the errors
resulting from the sampling of the spectrum with a resolution smaller than 
what corresponds to the characteristic line widths. This investigation has been made 
by making a number of sets of models with 30,000 wavelength points
which are identical except for the set of wavelength points used. 
We then find the temperature errors for $\tau_{\rm Ross} \ge 10^{-4}$
to be less than 0.003$\times T_{\rm eff}$ K with the maximum errors in the upper 
layers of the model (above $\tau_{\rm Ross}=0.01$). The corresponding errors in the 
flux integrals and flux derivative integrals (i.e. the integrals in Eqs.(\ref{eq27}) and (\ref{eq29}) 
by the sampling procedure are less than 1.5\%.  For the grid models, with
108,000 wavelength points, the sampling errors are correspondingly smaller
(by approximately a factor of $(108,000/30,000)^{-1/2}$). They then lead to
temperature errors of about 0.002$\times T_{\rm eff}$ K or less. 
For the M and C star models, we have found the corresponding errors in fluxes and
temperatures to be reduced by a factor of three or more. This is a result of the 
overlapping molecular-line absorption which reduces the dominance of a few
sampling-demanding strong metal lines in the flux derivative integrals in the 
upper model layers. 

It should be noted that
the fluxes, even if they are monochromatically precise within the model
constraints, may be far from representative for the regions surrounding each 
wavelength point in the spectrum. Thus, model fluxes integrated to represent
narrow spectral regions, $\Delta \lambda$  wide, will have relative errors
of typically $0.3\times[\lambda/(R \Delta \lambda)]^{1/2}$. E.g., for 100\AA  wide bands
at 5000\AA, the sampling errors will correspond to standard deviations of typically 1.6\%.
Obviously, synthetic spectra more detailed than the model fluxes must be calculated
for theoretical calibrations of e.g. narrow-band photometry. We have included the surface
fluxes in all 108,000 wavelength points into the model files 
published on {\sl http://marcs.astro.uu.se} -- more detailed high-resolution synthetic 
spectra will also be gradually added.  

For the integration of the $\mu$ moments of the radiative intensity (in practice 
$j_\lambda$) we use the $\mu$ points defined by the angles in the crossing points between 
the different parallel rays with impact parameter $p_i$ and the concentric spheres
around the stellar centre with radii $r_k$ corresponding to the discretization of the 
vertical $\tau_{\rm Ross}$ scale, $\tau_k$ cf. Fig 1. 
I.e., $\mu_{ik} = (r_{k}^2 - p_{i}^2)^{1/2}/r_k$, 
directions for which  $j_\lambda(\mu)$ is directly available. Considering the 
character of the radition field in a spherical atmosphere this is an adequate 
set of $\mu$ points, in spite of the fact that they are different for different 
radii, and different wavelenghts.  The typical number of $\mu$ points is 6 in
the plane-parallel case and typically 40-60 in the spherical models.
The quadrature in $\mu$ is performed using piece-wise cubic splines, as decribed 
by Nordlund (1984). 

\subsection{Representation of line opacity}

The line opacities have been pretabulated in the following way: For each 
individual species (e.g. H\,{\sc i}, C\,{\sc i}, C\,{\sc ii}, CH, C$_2$ etc.) and the full
set of wavelength 
points, arrays of line absorption were calculated for a number of $T-P_6$ points 
(or $T$ points for the molecules).
P$_6$ is given by
\begin{eqnarray}
\label{eq31}
P_6 = P({\rm H\,I}) + 0.42 P({\rm He\,I}) + 0.85 P({\rm H}_2)
\end{eqnarray}
where the static dipole polarizabilities and mean velocities of He\,{\sc i} and H$_2$
relative to that of H\,{\sc i}
have been used to approximately scale the broadening for H\,{\sc i} to these species.
Altogether, 17 $T$ values and 18 $P_6$ values were chosen.  Such tables were 
contructed for 6 different values of the microturbulence parameter, $\xi_{\rm t}
=0$, 1, 2, 5, 10 and 20 km/s.
Next, for a given chemical composition and microturbulence parameter 
the tables were summed for 
all atoms and ions, over the different species, with due consideration to the
ionization equlibria. This gave one table of the summed atomic
line absorption at each lambda point for each $T$, $P_6$ and $xi_{\rm t}$ chosen.
For each molecular species a table was produced for a combination of
12 $T$ and 6 $\xi_{\rm t}$ values.
In the subsequent model-atmosphere calculations, the logarithmic line absorption
coefficients were found from these tables by spline interpolation
to the appropriate temperatures and logarithmic pressures.
If models were needed at microturbulence parameters different from 
those of the tables, also interpolation can made. 

In the calculation of opacity data tables
the relevant absorption for each spectral line
at each $T$ and $P_6$ value was added to the OS table for wavelengths
points progressing to the blue and the red side of the central line wavelength
as far as the line opacity exceeds a pre-defined cross section. Also for very weak lines each line
was then included at at least two wavelength points. 
To limit the total computing time needed and still guarantee that no significant opacity be lost, 
these limiting cross sections were empirically determined for groups of
atoms and ions.

\subsection{Computing time, starting model and convergence}

The total computing time needed for the calculation of a model
atmosphere (the pre-calculation of the absorption coefficient tables 
excluded) scales approximately as follows: (1) For the radiative transfer 
part of the problem the time is proportional to $n_{\mu}\times n_{\tau}\times n_{\lambda}$.
(2) For the calculation of ionization equilibria and molecular equilibria the time scales as
$n_{\tau}$ and (3) for the calculation of absorption coefficients as $n_{\tau}\times n_{\lambda}$.
In addition to that, (4) some initiation time is needed. Setting $N_{\tau}= n_{\tau}/28$,
$N_{\lambda}=n_{\lambda}/10^5$ and $N_{\mu}=n_{\mu}/6$ and noting that we usually set $n_{\mu}=6$ in 
the plane-parallel case and that $n_{\mu}\approx 2/3\times n_{\tau}$ in the spherical case, we find
empirically that the time $t_{it}$ needed per iteration is, in seconds,  
\begin{eqnarray}
\label{eq41}
t_{it}\approx 10 + 10 N_{\tau} + 5 N_{\lambda} + 10 N_{\tau}N_{\lambda} + 2 N_{\tau}N_{\lambda}N_{\mu}.
\end{eqnarray}
For a typical spherical model with 108,000 wavelength points about 80 seconds per iteration are needed 
on a Mac Pro quad Intel Xeon 2.66GHz computer for one processor. 
For a plane-parallel model this is reduced to about
60 seconds. The relative small reduction reflects the fact that interpolation in the line-absorption
tables (the $N_{\tau}N_{\lambda}$ term) constitutes a major fraction of the computing time; only for
a highly spherical case (when $N_{\mu}$ becomes great) the radiative transfer calculations
dominate. No doubt, the calculations of absorption coefficients and their derivatives
could be speeded up further, e.g. by pre-tabulation. This would primarily be of interest at the
calculation of extensive model grids, or of models with more complex physics than decribed by the 
approximations in Sec. 3, e.g. as regards hydrodynamics.  

In multi-dimensional Newton-Raphson schemes like the present one,
convergence is rapid, provided that a starting solution which is
close enough to the final solution has been chosen. 

When model calculations are started from scratch with the MARCS program, 
usually
a gray starting model is chosen for the radiative zone. When the 
calculated flux in the convectively unstable zone of the first starting model exceeds the total flux, the 
temperature gradient is automatically and directly reduced, beginning at the onset of 
the convective instability, until the convective flux is smaller
than the prescribed total flux. In the present
grid, we usually start from a nearby model in the fundamental-parameter
space. If the starting model has a different effective temperature, a 
simple scaling of the temperature structure $T(\tau_{\rm Ross})$ may be applied,
but this is not necessary if steps of only a few hundred K are taken in $T_{\rm eff}$. 

The route towards convergence is often rapid but not always quadratical.
The pronounced non-linearities, not the least in the temperature
dependencies of the molecular equilibria with strong effects
on the opacities may slow down the convergence if one is not very close
to the final solution. Also, the changing 
presence and depth of the convective zone with temperature and metallicity
affect the convergence. 
Usually, convergence to temperature corrections althrough the models of
less than a few K, and logarithmic pressure corrections smaller than 
0.01 dex, is obtained after 4 to 10 iterations. For the models
with $T_{\rm eff} \ge 4000$ K the convergence is close to quadratic with
corrections reduced by almost one order of magnitude from one iteration to the next.
For the cooler models the convergence is slower but still fast. For some
parameter choices, however, converged models were not obtained. This
is in particular the case for the models with a strong radiative pressure
gradient (i.e. close to the effective Eddington limit, see Gustafsson \&
Plez 1992) which is close
to upsetting the hydrostatic equilibrium condition, i.e. models with high
temperatures and low gravities. Another difficulty occurs for models in a 
small band in the $T_{\rm eff}-{\rm log} g$ diagram, extending from about $T_{\rm eff} =8000K$, 
log$g = 5$, to
$T_{\rm eff}=6750K$, log$g = 3$, where convergence is not achieved in the deepest layers
of the models because the convectively unstable regions are swapping between thin 
convective zones and zones extending to 
depths below the bottom of the model. This difficulty may be circumvented
by extending the depth scale of the model to deeper layers. Some of the models, though not
fully converged in the deepest layers, are nevertheless presented in the grid
since the spectrum forming regions (above $\tau_{\rm Ross} = 10$) are not affected
at all. Convergence problems occurred for some of the coolest models with H$_2$ convection
zones at the surface as well as for models where the radiative pressure totally dominates the gas pressure. 
   
\section{General properties of the models}
\begin{figure}
\resizebox{\hsize}{!}{\includegraphics{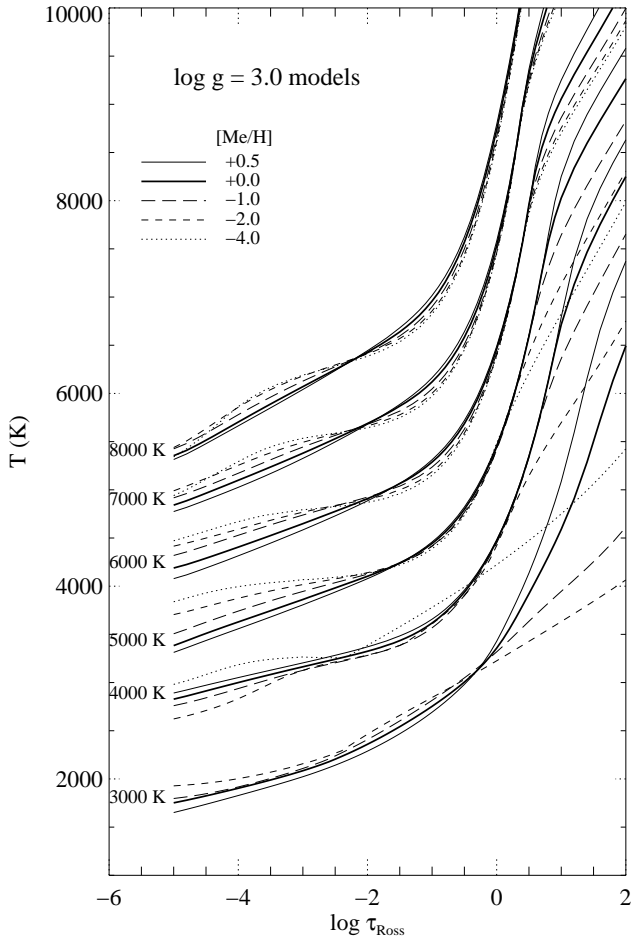}}
\vspace{5mm}
\caption{
The temperature structures for a set of model atmospheres with 
different $T_{\rm eff}$, log$g = 3$ and different metallicities.} 
\label{TTau}
\end{figure}

All models of the grid are accessible via {\it http://marcs.astro.uu.se/}. The organization of the model files is 
described there. Details of structures, thermodynamic variables,
molecular partial pressures and fluxes in $10^{5}$ wavelength points are given in the files. 
In Fig.\,\ref{TTau}
we present some sample temperature structures of grid models with different effective temperatures and metallicities. 
It is seen that the variation with metallicity at the surface and in the deep layers
increases when $T_{\rm eff}$ decreases from 8000 \,K 
to 5000\,K, and that this
variation with metallicity changes sign at 4000\,K in a non-trivial way. 
Subsequently, this and other properties of the grid models will be commented on. Our discussion here
is confined to general properties of the model structures, while more details on models of particular
types of stars, as well as discussion of model fluxes are deferred to later papers in this
series. 

\subsection{General effects of blanketing}
The effects of spectral lines on stellar atmospheres ("blanketing") have been explored thoroughly since it became 
possible to calculate models with lines included in the 1960s and 1970s. Here, we shall make a short summary 
of these results as a background for some more specific comments on the blanketing effects in the new grid.

In general, the blocking of
radiation leads to heating of the deeper layers of the atmospheres ("back-warming"). Here and subsequently we measure the
back-warming effect at a given continuum optical depth, in practice at $\lambda=500$ nm (i.e. {\it not} $\tau_{\rm Ross}$ 
which for the cooler models is strongly
affected by line absorption). The back-warming is in fact
easily estimated, directly from the definition of the effective temperature,
by assuming that a model with effective temperature $T_{\rm eff}$ is heated such that the deeper layers 
correspond to an unblanketed model with an effective temperature, 
\begin{eqnarray}
\label{eq411}
T_{\rm eff}'=(1-X)^{-{1\over 4}}\cdot T_{\rm eff}\,,
\end{eqnarray}  
where $X$ is the fraction of the integrated continuous
flux blocked out by spectral lines,
\begin{eqnarray}
\label{eq4111}
X={\int_0^\infty (F_{cont}-F_{\lambda}) d\lambda \over \int_0^\infty F_{cont} d\lambda}.
\end{eqnarray}  
The validity of this relation has been tested for models with different 
fundamental parameters by calculating X of these models, and next unblanketed models with effective temperatures 
revised according to the relation were constructed. Next, these models were then compared with the corresponding blanketed ones. The test came out 
favourably -- for models all over the parameter space of the grid, the $T_{\tau_{\rm Ross}}$ structures in the
interval $-1\le \tau_{\rm Ross}\le 0.5$ were reproduced astonishingly well by the corresponding unblanketed ones;
the additional steepening of the temperature gradient accross the $\tau_{\rm Ross}$ interval for the 
blanketed model with solar abundances may typically amount to $0.02\cdot T_{\rm eff}$ per decade in $\tau_{\rm Ross}$.

\begin{figure}
\resizebox{\hsize}{!}{\includegraphics{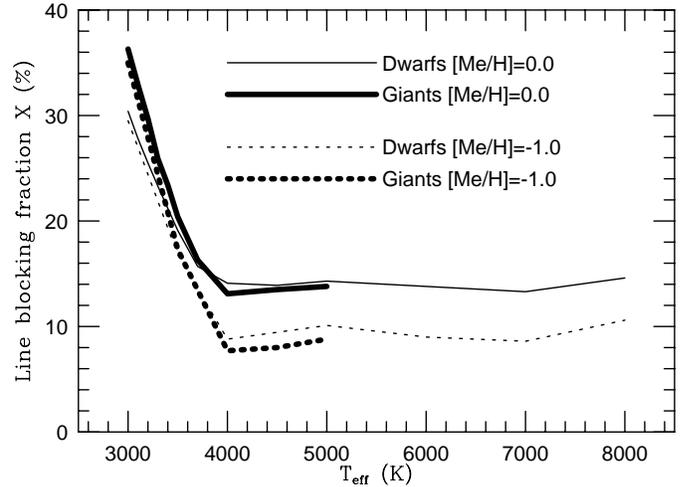}}
\caption{The blocking fraction $X$ in percent for models in the grid with two different metallicities. 
The dwarf models all have log$\,g=4.5$ while the giant models 
have log$\,g$ values increasing with temperature, from log$\,g=0.0$ at $T_{\rm eff}=3000\,$K to log$\,g=3.0$ at $T_{\rm eff}=5000\,$K.}
\label{Blockfrac}
\end{figure}

We have also calculated the total blocking fraction $X$ 
for the grid models and find the remarkable behaviour shown in Fig.$\,$\ref{Blockfrac}.
Thus, $X$ stays nearly constant for the temperature interval $8000\,{\rm K}\ge T_{\rm eff} \ge 4000\,$K for a given metallicity, and
is closely identical for dwarfs and giants, while below 4000\,K it increases as temperature goes down, due to the heavy TiO and H$_2$O
blocking. Here, the metallicity sensitivity of $X$ decreseas (which reflects the fact that the molecular lines fill in the
continuum regions in the spectrum and define the $\tau_{\rm Ross}$ scale; $T(\tau_{\rm Ross})$ being rather robust against metallicity
changes) while the gravity sensitivity increases. 

\begin{figure}
\resizebox{\hsize}{!}{\includegraphics{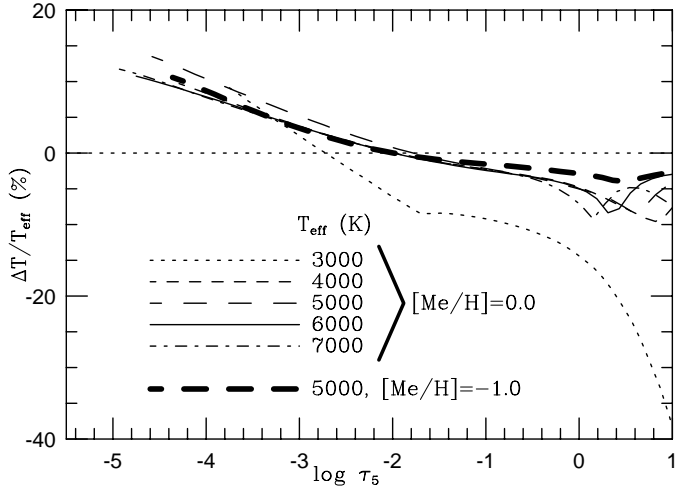}}
\caption{The effects of switching off line absorption on the temperature structure of a sequence of models with log$g=3.0$ and solar
metallicity. Note that $\Delta T \equiv T({\rm no lines})-T({\rm lines})$. It is seen that the blanketing effects are fairly independent of effective temperature for models
with $T_{\rm eff}\ge 4000.$}
\label{blankeff}
\end{figure}

For the upper layers of the atmospheres the situation is more complex,
as was discussed e.g. by Gustafsson et al. (1975), Gustafsson \& Olander (1979) and Carbon (1979) and as is illustrated
in Fig.\,\ref{blankeff}.

In short, for LTE the spectral-line absorption may cause a cooling or a heating, depending on whether the absorption is located
on the long-wavelength side of the local Planck function (cooling, Case 1) or on the short-wavelength side. In the latter case,
heating (2a) or cooling (2b) may result, depending on whether the absorption is concentrated to the upper layers (2a) or 
whether it extends to the deeper atmosphere (2b). Examples of Case 1 are the cooling by the IR VR lines of CO for models of late G stars 
(cf. Johnson 1973, Gustafsson et al. 1975) and by H$_2$O in the IR for later M stars. A clear example of Case 2a is the heating by
the electronic bands of TiO in early M-star models (Krupp, Collins and Johnson 1978) while a Case 2b example is provided by the metal-line
absorption in the UV and blue for F and G stars. We note in passing that these surface effects are all dependent on the LTE assumption -- in
particular if the spectral lines are partially formed in scattering processes the effects may be significantly smaller due to the weaker coupling
of the radiation to the temperature of the local gas. The backwarming effects, on the other hand, are not strongly dependent on the 
assumed line-formation mechanism as long as the amount of blocking is unchanged. Missing or erroneous 
line absorption data may, however, still be of significance as a source of systematic errors in the back-warming effects. 

The general effects of blanketing are illustrated for the grid models by a test where all line absorption 
was swichted off. All models with $T_{\rm eff} \ge 4000 \, K$ and solar metallicity 
show a similar reponse: 
the temperature is increased by about $10 \,\%$ in the outermost layers if the line opacity is neglected, 
an increase which gradually diminishes to
zero at $\tau_{{\rm Ross}}=0.01$ and then changes to a lowering of the temperatures at depth since back-warming 
is inhibited. 
(cf. Fig.\,\ref{blankeff}). For the coolest giants
the effects become considerably greater. 

We shall now give some more specific comments 
on effects of blanketing in the new grid, with remarks on the 
effects of various opacity sources, and on the effects of changing abundances and microturbulence. Furthermore, we shall
comment on the effects of sphericity, and explore some interesting coupling between these latter effects and blanketing.

\subsection{Effects of different opacity sources}
\begin{figure}
\resizebox{\hsize}{!}{\includegraphics{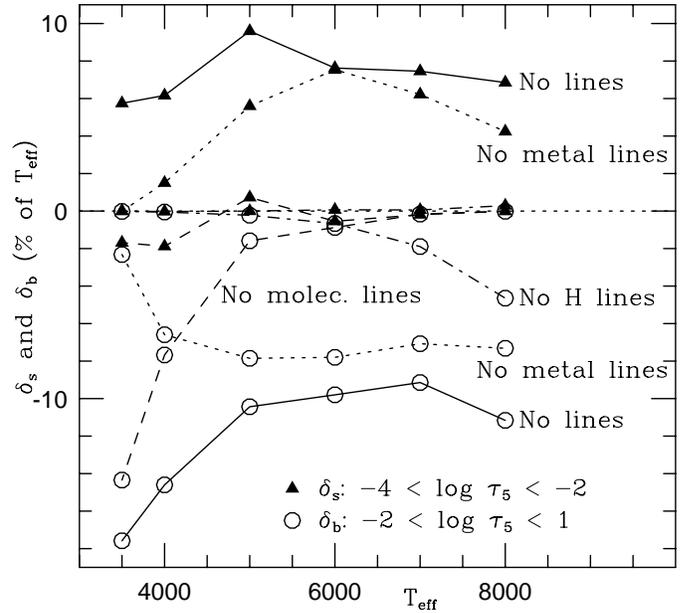}}
\caption{Structural effects on the surface (black triangles) and depth (open
circles) of model atmospheres by different groups of spectral lines.
The solid lines show 
surface-heating and bottom-cooling as defined in Eq.\,(\ref{eq412}) when
all spectral lines are removed in the calculation of
solar-metallicity dwarf models ($\log g=4.5$).
The dotted, dashed, and dash-dotted lines show, respectively, the
effects of removing only the metal lines, the molecular lines, or
the hydrogen lines.
The structural effects of spectral lines are surprisingly similar
for all solar-type stars.
The removal of all line opacity makes the models quite
isothermal in the optically thin surface layers.
}
\label{HeatCool}
\end{figure}

In Fig.\,\ref{HeatCool}
we present the effects of different opacity sources, measured as integrated effects on the surface and at depth
in a set of models for main-sequence stars. We have thus defined the quantities
\begin{eqnarray}
\label{eq412}
\delta_s \equiv \int_{-4}^{-2} \Delta T(\tau_{5}) d{\rm log}\tau_5 \,\, / \int_{-4}^{-2} d{\rm log}\tau_5 \nonumber \\
\noindent \delta_b \equiv \int_{-2}^{1} \Delta T(\tau_{5}) d{\rm log}\tau_5 \,\,/ \int_{-2}^{1} d{\rm log}\tau_5.
\end{eqnarray}
$\Delta T$ is the quantity
\begin{eqnarray}
\label{eq413}
\Delta T(\tau_5) \equiv [T(\tau_5) - T'(\tau_5)]/T_{\rm eff}
\end{eqnarray}
where $T(\tau_5)$ is the temperature structure measured at the continuum optical depth at 5000\,\AA\,\, and $T'(\tau_5)$ the 
corresponding quantity for a modified model atmosphere with the same fundamental parameters but for which 
line opacity sources have been neglected one at a time or all simulataneously (such that the corresponding model is
unblanketed). The reason why $\tau_5$ and not $\tau_{\rm Ross}$ were chosen here is that the Rosseland mean is
directly affected by the spectral lines, in particular for the cooler stars, while we here wish to 
separate these effects on the temperature structure from those on the $\tau$ scale.  

We see from Fig.\,\ref{HeatCool} that for a sequence of main-sequence models with solar metallicity the total blanketing
effect, both at the surface and in the deeper layers, is rather temperature independent. Thus, the mean surface cooling
amounts to about $7\%$ of $T_{\rm eff}$ and the mean heating of the layers below $\tau_5=10^{-2}$ 
is about $11\%$, over most of the effective-temperature interval. I.e., the somewhat decreasing
importance of metal lines when proceeding towards the hot end of our effective-temperature interval is rather well compensated for
by the increasing hydrogen-line blanketing, and the diminishing effects of metals for the cooler models are compensated for by
the molecular blanketing. These compensation effects explain why relatively old differential abundance analyses, made by 
primitive model atmospheres such as scaled solar models, are often found to agree rather well with more recent results 
based on models with much more complete atomic and molecular data.
\begin{figure}
\resizebox{\hsize}{!}{\includegraphics{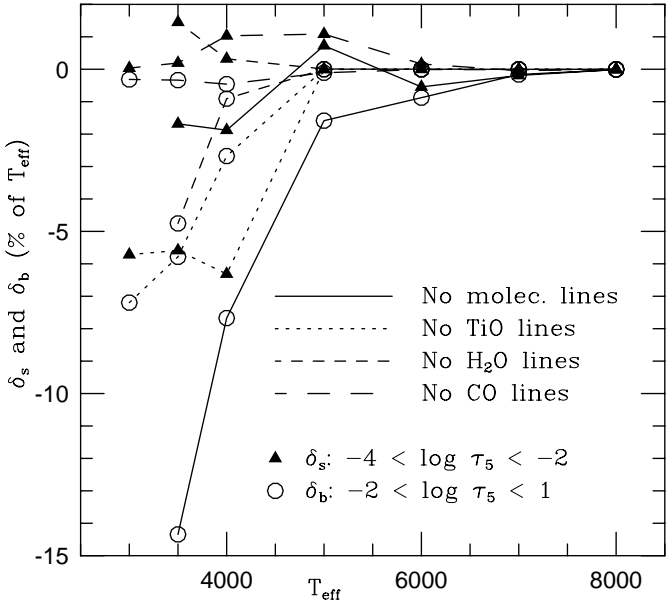}}
\caption{Similar to Fig.\ref{HeatCool}.
The structural effects of molecular lines separated for different species to show the
results of removing the line opacity of TiO, water, and CO, respectively.
}
\label{HeatCoolMol}
\end{figure}
In Fig.\,\ref{HeatCoolMol} we display the corresponding effects when 
absorption by different molecular species is left out. It is seen that the only significant molecular effects at the surface 
for these models are the surface cooling due to CO for moderately cool models and to H$_2$O for the coolest ones and the surface heating due
to TiO, while in the deeper layers the significant back-warming is due to TiO and H$_2$O. Other molecules, like CN, play only
a very minor role in this respect. 
      
\subsection{Effects of abundance changes} 
The variation of certain abundances in the MARCS grid lead to effects on the models. 
Such effects were explored for the CNO abundances of G and K giant models by Gustafsson et al. (1975), for M and S stars by
Plez et al. (2003) and for carbon stars by Lambert et al. (1986) and J\o rgensen, Johnson \& Nordlund (1992). The results given here are
illustrative and complementary to those of previous studies -- more detailed discussion of abundance effects will be presented
in later papers in this series. 

The recent considerable downward corrections of solar CNO abundances by Asplund and collaborators (Grevesse, Asplund
$\&$ Sauval 2007 and references therein) are found to lead to very minor effects for most regions in the
parameter space spanned by the present grid. Thus, for the temperatures at given $\tau_{{\rm Ross}}$,
the effects are less
than $1\,\%\times T_{{\rm eff}}$ for all depths and luminosities for giants at solar metallicity. The most pronounced effects in 
the surface layers appear in models with $T_{\rm eff}\le 3500\,$K and models with $T_{\rm eff}\approx 5000\,$K 
where the reduced 
cooling due to H$_2$O and CO, respectively, is visible. 
%For models of dwarfs with $T_{{\rm eff}} \le 4000\,{\rm K}$ the decreased blocking may noticeably change the
%back-warming in deeper layers, which then couples to a different structure in the convective zone. {\it MORE TO SAY HERE?}. 

\begin{figure}
\resizebox{\hsize}{!}{\includegraphics{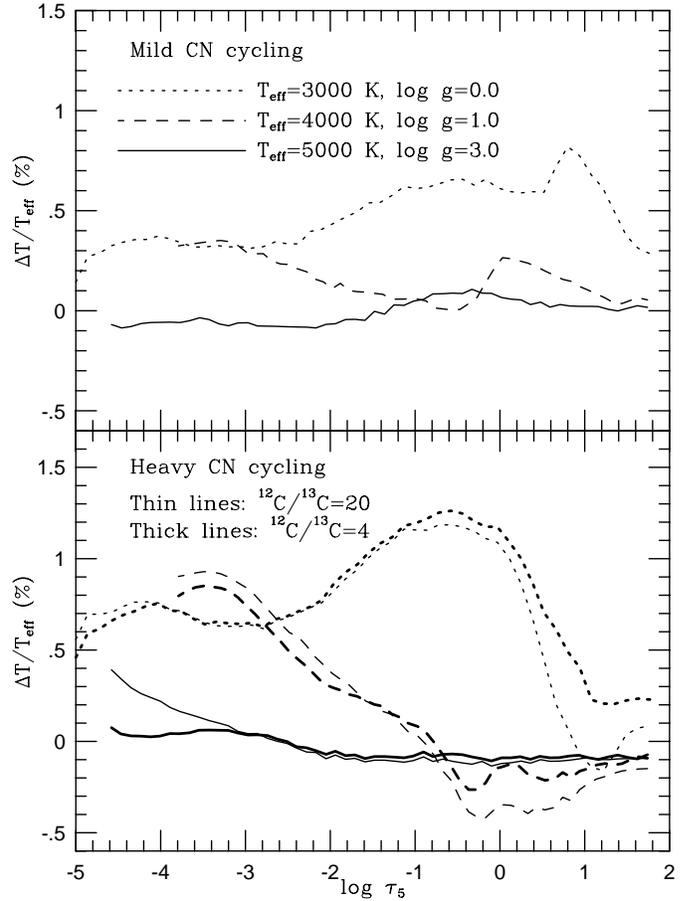}}
\caption{The effects of dredge-up of CN-cycled material to the surfaces of
Pop I giant stars as illustrated by model atmospheres with different C and N abundances.
In the top panel carbon has been converted to nitrogen such that
the C/N abundance ratio has been lowered from the solar ratio of
4.07 to 1.5, typical for the first dredge-up in low-mass stars. In the lower panel the ratio has been further lowered to 0.5
and there also the effects of changing the $^{12}$C$/^{13}$C
ratio are displayed. 
The stronger back-warming of the 3000\,K models is due to the increased
importance of water as more oxygen gets available when the CO abundance is decreased by the reduction of carbon while the
heating effect at the surface merely reflects the inadequacy of $\tau_5$ for this heavy line-blocked model -- for 
a plot of the temperature vs. $\tau_{\rm Ross}$, an increased cooling at the surface appears.  
In the 5000\,K model the decreased CO cooling leads to a hotter surface. 
}
\label{CNcycl}
\end{figure}
The CNO abundances in stars are modified through the first dredge-up along
the subgiant-giant branch, mostly converting C to N (see, e.g.,  Boothroyd \& Sackmann 1999, Charbonnel 1994). 
The effects of this on the model atmospheres 
are, however, small or moderate, as illustrated in Fig.\,\ref{CNcycl}.
The two most important effects are again the
enhanced H$_2$O surface cooling for the coolest models (since more O is available if C and thus CO is depleted), and a diminished
CO cooling for the somewhat hotter models. The changes due to effects by CN are marginal.
%It is seen that the total effects on the temperature structures are rarely greater than $1\,\%$ of $T_{\rm eff}$. Another 
%effect of CNO cycling of 
%some importance, illustrated in Fig.\,\ref{CNcycl} is
%due to 
The accompanying changes of $^{12}{\rm C}/^{13}{\rm C}$ ratios
%; the correspondingly increased line absorption from the isotopic lines 
are obviously only of minor significance.
% and for normal G and K giants 
%mainly due to an increased back-warming by $^{13}$CH, $^{13}$C$^{12}$C,
%$^{13}$CN and $^{13}$CO lines which partially compensates for the reduced back-warming from C molecules as the carbon abundance 
%is lowered.       

\begin{figure}
\resizebox{\hsize}{!}{\includegraphics{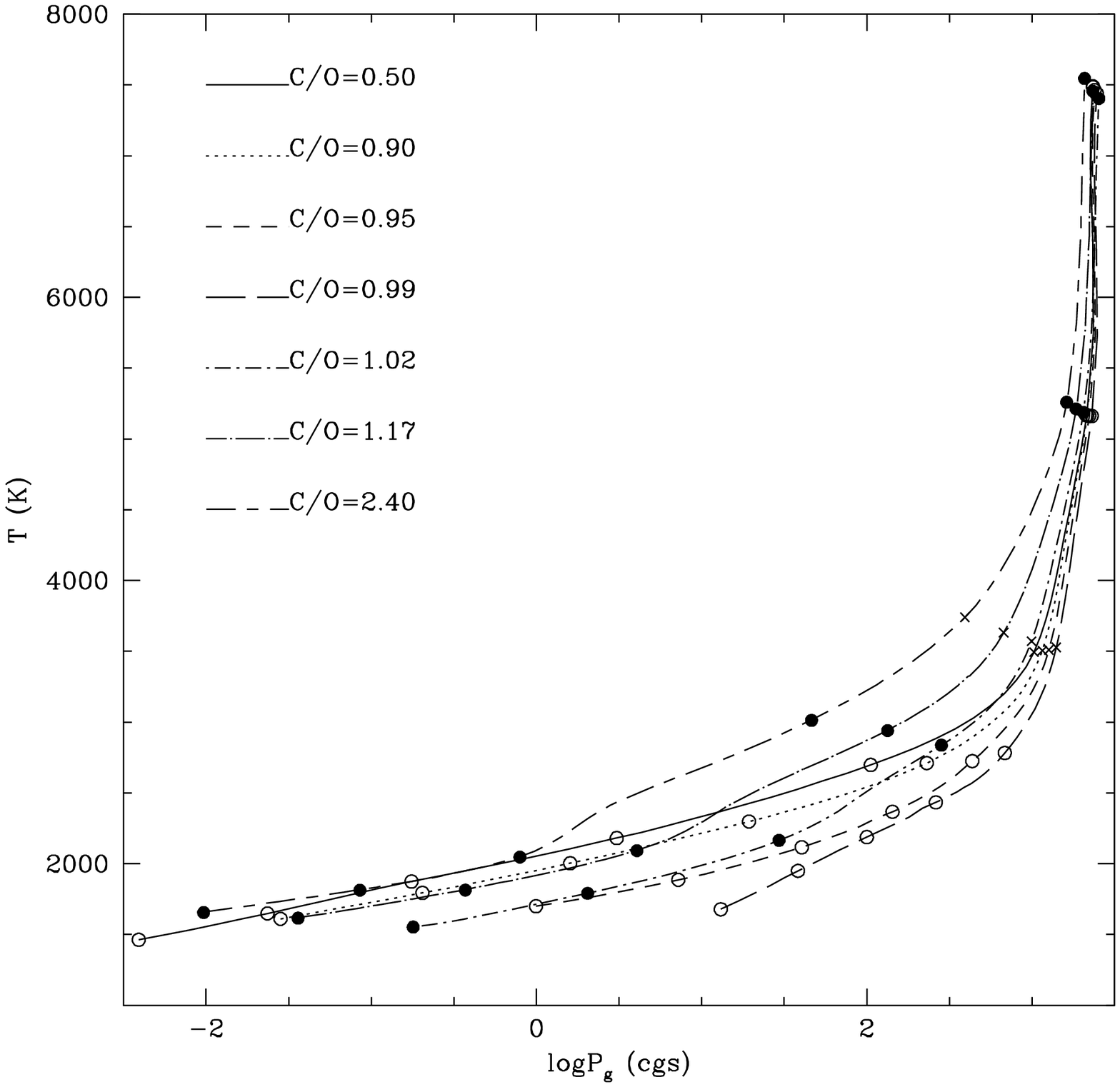}}
\caption{
A sequence of MARCS spherical model atmospheres with $T_{\rm eff}$ = 3000\,K, log$g$=0.0, $M=1\,M_{\odot}$ . 
The models have different abundance ratios, C/O, marked in the figure, but else solar metal abundances.
Points on the temperature-pressure relations with log$\tau_{\rm Ross}=-5$,
-4,-3,-2,-1,1 and 2 are marked with circles, which are filled for the carbon-rich models (C/O$\ge 1.00$).
Points where log$\tau_{\rm Ross}=0$ are marked with $\times$.
The figure demonstrates the strongly increasing pressures at given temperature and optical depth as the C/O
ratio increases, until C/O reaches 1.0 when the situation is drastically reversed as the opacities increase 
and the pressures decrease again.  
The models with C/O $\le 0.9$ show M-type model spectra, 
those in the interval $0.9 \le 0.98$ S-type spectra, while those with greater C/O show N-star spectra.}
\label{CarbonSequence}
\end{figure}

The most severe atmospheric effects of CNO abundance changes occur, however, when the carbon abundance is raised by dredge-up
on the AGB and, as the carbon abundance approaches or exceeds that of oxygen, the star becomes an S or a C star, respectively,
with drastic changes in radiation fields and thus temperature-pressure structure. 
This transition has been explored using MARCS models
(for references, see above), and will be further discussed in Paper V and Paper VI in the current series. As an illustration
here, however, 
we present a sequence of models, with C/O ratios ranging from 0.5 to 2.4 in Fig.\,\ref{CarbonSequence}.
  
\begin{figure}
\resizebox{\hsize}{!}{\includegraphics{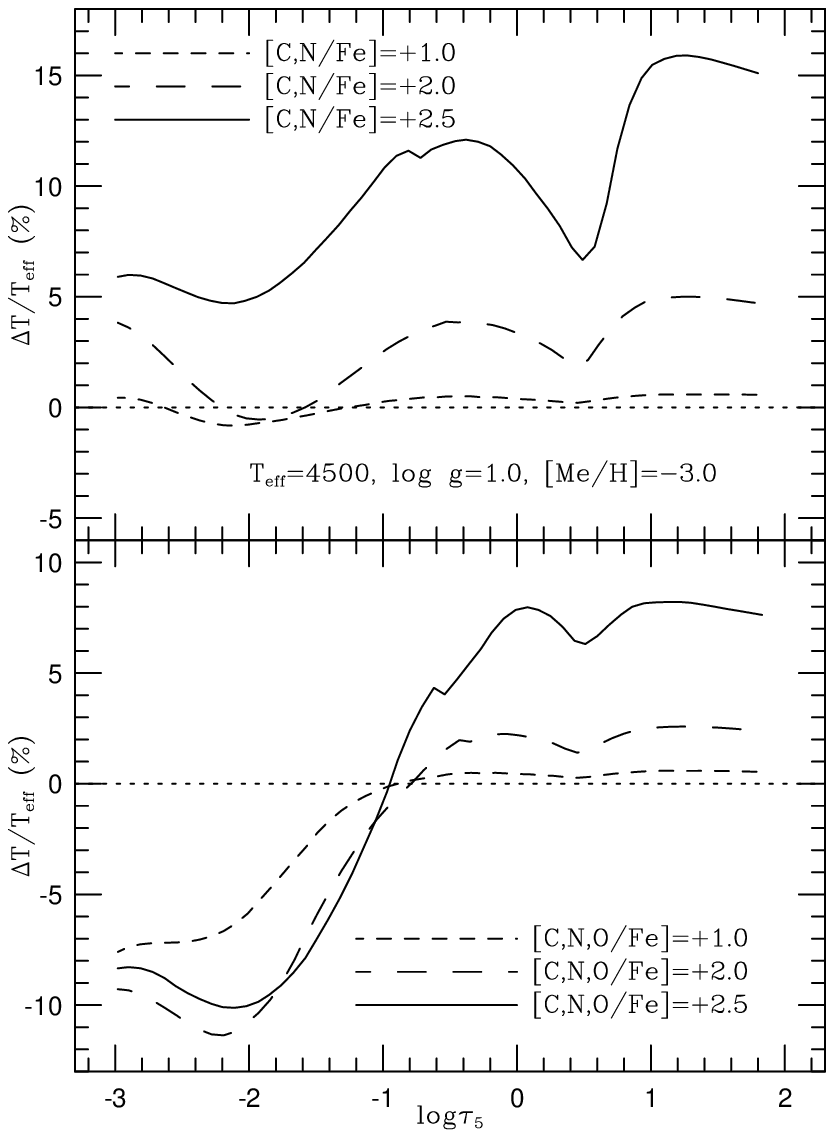}}
\caption{
The effects of non-solar CNO abundance ratios on extreme Pop II star
models with $T_{\rm eff}$=4500\,K, log$g=1.0$, [Me/H]=$-$3.0.
In the top panel the carbon and nitrogen abundances have been
increased by factors of 10, 100, and 300 above the solar ratio
to iron, oxygen and all other elements. The heating in the
middle layers, primarily due to C$_2$, is strong, and even more so
at depth which is due to the increased electron pressures and 
continuous opacities. 
In the lower panel also oxygen is enhanced.
The increased electron pressures in the more CN(O)-rich models 
lowers the density and
convection efficiency at a certain optical depth which
increases the temperature gradient.
The major surface cooling agent in the outer surface layers is CO.
}
\label{Pop3CN}
\end{figure}

The most metal-poor stars known until now show severe departures from "standard" CNO abundances: 
they seem to be comparatively very rich in the CNO elements (cf. Christlieb et al. 2004, Frebel et al. 2005, 
Norris et al. 2007 and references therein). 
In general, there is a group of low-metallicity
stars with considerably enriched carbon abundances, so called Carbon-Enriched Metal-Poor (CEMP) stars, (see, e.g., Aoki et al. 2007), which
may show carbon enrichments ranging in the interval between a factor of 10 and 1000 relative to a solar C/Fe, as well as
similarly considerable N enrichments and possibly also O enrichments. 
The effects on the atmospheric structures from these enrichments are illustrated in Fig.\,\ref{Pop3CN} for one representative 
CEMP giant model.

It is seen that the enhanced C and N abundances heat the layers $-2\le {\rm log}\tau_{5}\le 0.5 $ 
considerably for these model parameters (a result which was also found
by Hill et al. 2000). Also the deep layers get warmer, which is due to the increased electron contributions
from the enhanced carbon. This increases the H$^{-}$ opacity which lowers the density at a given optical
depth and therefore reduces the convective energy flux which increases the inner temperature gradient. 
These effects change if also O is enhanced correspondingly. Then the CO cooling takes over in the surface layers, 
and much of the C$_2$ and CN
heating absorption vanishes, since most of the carbon is bound in the CO molecules. 

\begin{figure*}[ht]
\centering
\resizebox{\hsize}{!}{\includegraphics{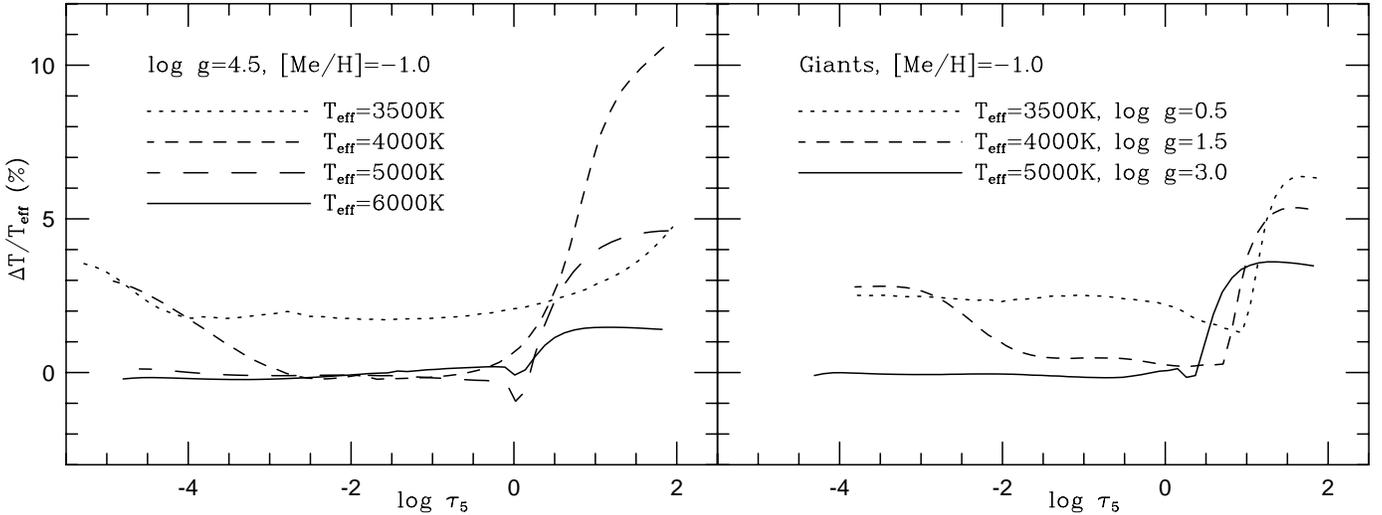}}
\caption[]{The effects of an increase in the $\alpha$-element abundances by 
+0.4\,\dex on the temperature structure of models of dwarfs and giants with [Fe/H]=-1.0.
The comparison model in each case is a corresponding model with no $\alpha$-element enhancement ([$\alpha$/Fe]=0.0).
The increased electron pressure lowers the density and
convection efficiency at a certain Rosseland optical depth which
increases the temperature gradient in the deeper parts of a model.
}
\label{Alpha}
\end{figure*}
The effects of the observed increase of $[\alpha/$Fe] for galactic, though seemingly not all, metal-poor stars can also
be explored using the grid models. Although these changes are easily seen in spectra of the models, e.g. in the strengths of the
Mg and Ca lines, the effects on the temperature-pressure structures are only minor. As is seen in Fig.\,\ref{Alpha} they are, however, of 
some significance for models with $T_{\rm eff}<4500 \,{\rm K}$ where the higher Ti abundance increases the surface heating
as well as the backwarming of the TiO absorption. Simultaneously, the electron contributions from Mg and Ca 
increase and raise the H$^-$ opacity, which partly reduces the effects of the Ti abundance change. 
As discussed above for the CEMP star models, the higher electron pressures also reduce the densities at a given optical depth, which 
diminishes the convective flux and thus admits a stronger temperature gradient in the deep atmosphere. 

It should finally be noted that systematic errors in the models, due to missing opacities, departures from LTE 
or convectively generated inhomogeneities and errors in calculated convective fluxes, may well vary systematically 
with chemical composition. Thus, the trends discussed in the present section may have to be revised when more realistic
models are available.

\subsection{Blanketing effects of turbulence}
In the present grid, there are models calculated with different microturbulence parameters $\xi_t$, but with otherwise
identical input data. This makes it possible to systematically explore the dependence of the blanketing effects
on microturbulence. Before discussing the results some considerations of what one could expect will be presented.

\begin{figure}
\resizebox{\hsize}{!}{\includegraphics{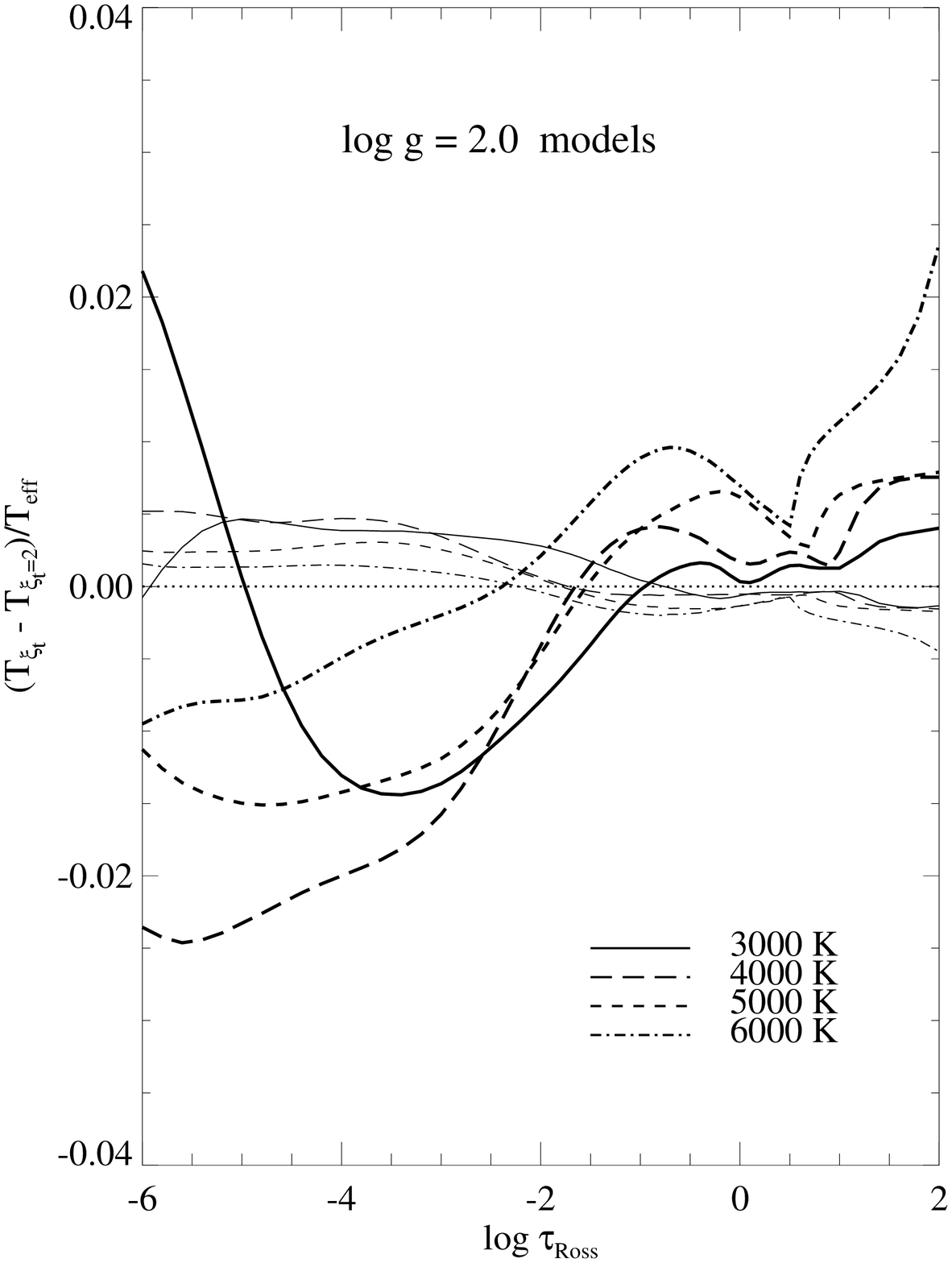}}
\caption{
$T(\tau_{\rm Ross})$ for models with different microturbulence parameters $\xi_{\rm t}$, minus $T(\tau_{\rm Ross})$ for a model with $\xi_{\rm t}=2$ km/s, for solar
composition, log\,$g$=2.0 and different effective temperatures. Thick lines denote models with $\xi_{\rm t}=10$ km/s, thin lines with $\xi_{\rm t}=1$ km/s.} 
\label{tdiffmic}
\end{figure}

An increased microturbulence enhances the blocking of the spectrum by isolated 
saturated spectral lines, in proportion to $\xi_t$. So, if all lines were saturated one might estimate from
Eq.\,(\ref{eq411}) that the backwarming $\Delta T/T$ would increase by about $1/4\cdot\Delta \xi_t/\xi_t\cdot X$. However, this is an upper limit
since a 
considerable fraction of the spectral lines (though not necessarily those that contribute most significantly 
to the blocking) will not be saturated -- their equivalent widths will be uneffected by a changing $\xi_t$ and thus the total blocking
not changed. Also, for more metal-rich stars the overlapping of strong lines reduces the differential
effects of a $\xi_t$ increase, as does the fact that many of these lines have strong damping wings. We thus
expect the effects of microturbulence changes on the backwarming to be small for the most metal-poor stars where
even the strongest lines are unsaturated, 
then increase but finally level off for the most metal-rich and coolest models, and never reach the 
estimated upper limit. These expectations are verified by the models, as is 
illustrated in Fig.\,\ref{tdiffmic}. The differential back-warming caused by an increased microturbulence parameter,
e.g. by a factor of 2, is of about the same magnitude as the result of doubling the metallicity. This suggests that the total blocking
contribution of lines on the flat part of the curve-of-growth (mainly sensitive to $\xi_t$) is of similar significance
as that of weak spectral lines (proportional in strength to the metallicity); this picture is complicated
by both the effects of lines on the damping-part of the curve-of-growth (with only a square-root dependence on abundance)
and the strong abundance sensitivity of some diatomic molecules. 
As regards the differential 
blanketing effects in the upper atmosphere when the microturbulence parameter is changed, the situation is not quite
obvious. The first, heating, part of the radiation equilibrium 
integral in Eq.\,(\ref{eq29}) will not change in itself if only the line-absorption
profiles become broader. If the effects
on $J_\lambda$ are taken into consideration, the decrease of the intensity over larger wavelength regions by the widening of the lines will, 
as long as they are weak, be compensated for by the increase of $J_\lambda$ in the line center. 
However, for saturated lines $J_\lambda$ will be reduced over broader wavelength intervals
when $\xi_t$ is increased, and thus the heating term will become smaller. The second, cooling, term
will not change from a broadening of the
line profiles. Thus, a net cooling is expected to occur. We see from Fig.\,\ref{tdiffmic} that in practice
increasing $\xi_t$ from 1 to 2 km/s leads to some moderate cooling (on the order of $0.5\,\%$ in $T/T_{\rm eff}$)
of the upper layers for all effective
temperatures while further increase up to 10 km/s leads to further cooling 
(by up to $2\%$, for the coolest models), however, 
this balance shifts to heating in the very surface layers.

Changes in the macroturbulent parameter (classically representing velocity shifts on geometrical 
scales corresponding to at least one optical depth unit in the continuum) are not able to change the model structures through
radiative field changes; concerning effects of changes in the turbulent pressure which may be related to both macro- and 
microturbulent motions, see Sec 3, above. 

Real stellar atmospheres are not believed to be well represented by the concepts of micro- and macroturbulence. Velocity 
gradients on intermediate scales are of vital importance, as has been shown by convective 
simulations (Nordlund \& Dravins, 1990, Asplund et al. 2000a), and global gradients caused by pulsations
as demonstrated in Lederer et al. (2006) and references therein. As regards the effects on the radiative field, these gradients 
allow for more heating in 
spectral lines since the gas high up in the atmosphere may absorb hot radiation from lower layers that was not absorbed below;
similarly, the gas may cool more efficiently through spectral lines since the shifts make the gas more transparent in overlying
layers. However, even more important are the hydrodynamical effects due to gas expansion and compression 
on the energy balance. Altogether, this makes the representation
of the energy balance in real stars by classical models quite problematic; also, there is no hope 
to gain further insight into the physics of these systems by fine-tuning the turbulence parameters.  
 
\subsection{Effects of sphericity}
The properties of spherically symmetric model atmospheres for late-type stars have been discussed or commented upon in
numerous studies. The area was first pioneered by Schmid-Burgk \& Scholz (1975) and Schmid-Burgk, Scholz \& Wehrse (1981 
and papers cited therein), 
and later explored by Plez (1990), Plez, Brett \& Nordlund
(1992), J\o rgensen, Johnson \& Nordlund (1992), Hauschildt et al. (1999) and Heiter \& Eriksson (2006). 
The papers by Plez and collaborators, by J\o rgensen and collaborators and by Heiter \& Eriksson, all
build on earlier versions of the MARCS code. These papers demonstrate that the effects of sphericity are 
in general important for the temperature structures, causing relative temperature effects on the order of 
1\% or more, for extensions of the atmosphere normalized on the stellar radius greater than about 2\%.
The relative extensions scale roughly as $g^{-1/2}$; the sphericity effects are thus
mainly important for low-gravity models.     
These papers do however not systematically
explore the effects on the models and fluxes of changing from plane-parallel to spherical symmetric
geometry. Here, some further comments will be made on this topic on the basis of the new
model grid. 

The {\it extension} $z_0$ of the stellar atmosphere, measured as the geometrical thickness of the
atmosphere (for instance defined as the layers with $-5.0\le {\rm log} \tau_{\rm Ross} \le 0.0$) is a key 
factor in the discussions. 
It is easy to estimate the extension from the hydrostatic-equilibrium 
equation, the definition of optical depth and the gas law for an ideal gas. The relative thickness of the 
atmosphere from optical depth $\tau$ up towards the surface at $\tau_0$ is thus: 
\begin{eqnarray}
\label{eq42}
{z_{0}\over R_1} = {{\cal R}<T>\over <\mu_{\rm mol}> g R_1(\eta+1)} \cdot {\rm ln} ({\tau\over \tau_0}) = 
4\cdot 10^{-7}{R_1^\odot \cdot T_{\rm eff}\over M^\odot\cdot (\eta+1)}.
\end{eqnarray}
Here, we have assumed the variation of the pressure in the atmosphere to be dominated by the density
variation -- the temperature and the mean molecular weight $\mu_{\rm mol}$ in the gas law were thus approximated by a
constant characteristic temperature and molecular weight, $<T>$ and $<\mu_{\rm mol}>$ (here set = 1.26), respectively. 
The mass absorption coefficient (opacity per gram) $\kappa_{g}$ is assumed to vary (only) with the density $\rho$, 
$\kappa_{\rm g}=k\cdot \rho^\eta$. 
%Another assumption made in
%deriving Eq.\,(\ref{eq42}) is that the mass absorption coefficient (opacity per gram) $\kappa_{g}$ does not vary with
%depth. This is not a very valid assumption -- opacities are pressure- and temperature sensitive and often
%increase inwards in the atmospheres, and for cooler stars also with height in the outer layers due to 
%molecule formation. If we describe this variation of $\kappa_{\rm g}$ as a dependence on the density $\rho$, 
%and write $\kappa_{\rm g}=k\cdot \rho^\eta$,
%Eq.\,(\ref{eq42}) is still valid if the rihgt-hand side is corrected by a factor $1/(\eta+1)$. We may then estimate the 
%relative extension $z_{0}/R_1$ where $z_{0}$ spans the $\tau$ interval $10^{-5}\le\tau_{\rm Ross}\le 1.0$:
%\begin{eqnarray}
%\label{eq43}
%{z_{0}\over R_1} = 11.6 {R_{\rm gas} T_{\rm eff}\over <\mu_{\rm mol}> g R_1(\eta+1)}=\\ \nonumber
%=4.0\cdot 10^{-7}{R_1^\odot \cdot T_{\rm eff}\over M^\odot\cdot (\eta+1)} .
%\end{eqnarray}
$R_1^\odot$ and $M^\odot$ are the model radius and mass, respectively, in solar units.

\begin{figure}
\resizebox{\hsize}{!}{\includegraphics{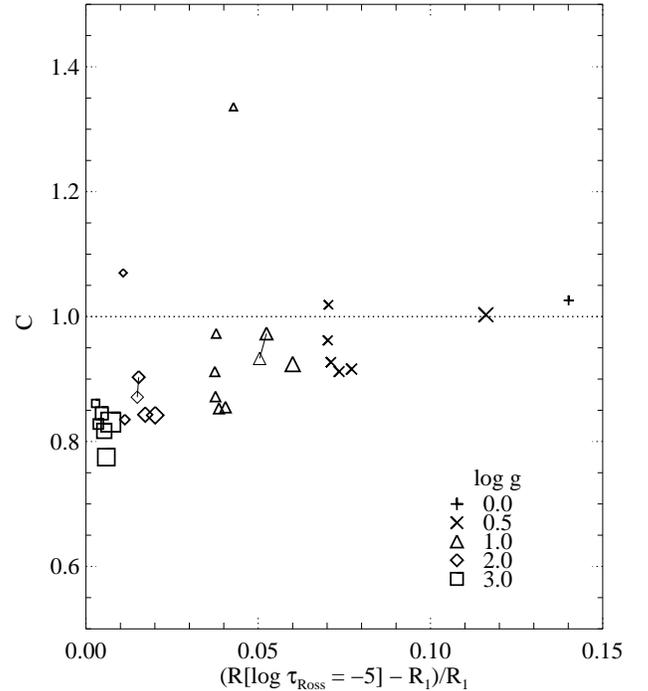}}
\\
\\
\caption{
The value of $C$ as calculated from Eq.\,(\ref{eq402}) with model values of $z_0/R_1$ for a set of models with
$T_{\rm eff}$ ranging from 3000\,K to 8000\,K (with bigger symbols the higher the effective temperature)
and metallicities [Me/H]=0.0, except for two models with [Me/H]$=-1.0$ which are 
attached with thin lines to corresponding [Me/H]=0.0 models. $\eta$ was chosen = 1. The value of $C$ when based on the value of the 
extension $z_0/R_1$ estimated from Eq.\,(\ref{eq42}) is 1.0 and shown by the horizontal dotted line.} 
\label{deltatradidiff}
\end{figure}

According to our analytical estimate of the relative extension in Eq.\,(\ref{eq42}), 
the quantity
\begin{eqnarray}
\label{eq402}
C=2.5\cdot 10^6[{z_0\over R_1}]\cdot {M^\odot \over R_1^\odot \cdot T_{\rm eff} } \cdot(\eta+1)
\end{eqnarray}
should be approximately constant and equal to 1.0. It is seen in Fig. \ref{deltatradidiff} that the
relative extension is fairly well accounted for by the approximate expression, and that the choice 
of $\eta=1$ is a reasonable fit
for an extensive part of the HR diagram, which is to be expected since the dominating H$^-$ absorption per gram
is roughly proportional to the pressure. Note that the relative extension of an atmosphere 
is relatively independent of the magnitude of the opacity, which cancels in the zero-order approximation. A 
depth-variation of $\kappa$ is of some significance, but only marginally so. The most
important sphericity effects related to the opacity are more indirect, through the effects that a changed temperature structure
may lead to, in particular when molecular absorption is switched on, as will be demonstrated below. 

We now turn to the sphericity effects on the temperature structure of the models. The temperature in radiative
equilibrium is set by the radiative field, following Eq.\,(\ref{eq29}). The basic effects due to sphericity at a 
point high up in a stellar atmosphere come
from the general dilution of the radiative field, partially due to the 
absence of incoming radiation from angles $\mu\le \mu_0$ where $\mu_0$ gradually increases the further from the 
optically thick layers the point is situated. Basically, this leads to less radiative heating of the 
upper layers of the atmosphere and thus cooler temperatures there. One may prove that this reduction of the mean intensity
for small extensions will scale with $(z_{0}/R_1)^{1/2}$.  
For larger distances from the star, a simple and approximative estimate for the temperature
effect, was presented by Gustafsson et al. (1975) (see also B\"ohm-Vitense 1972) who assumed that the radiative
flux should be roughly proportional to radius $r^{-2}$ and to temperature $T^4$ as one may estimate from 
Stefan-Boltzmann's law. This leads to the estimate 
\begin{eqnarray}
\label{eq403} 
\Delta T/T \approx -0.5\times z_0 /R_1. 
\end{eqnarray}

\begin{figure}
\resizebox{\hsize}{!}{\includegraphics{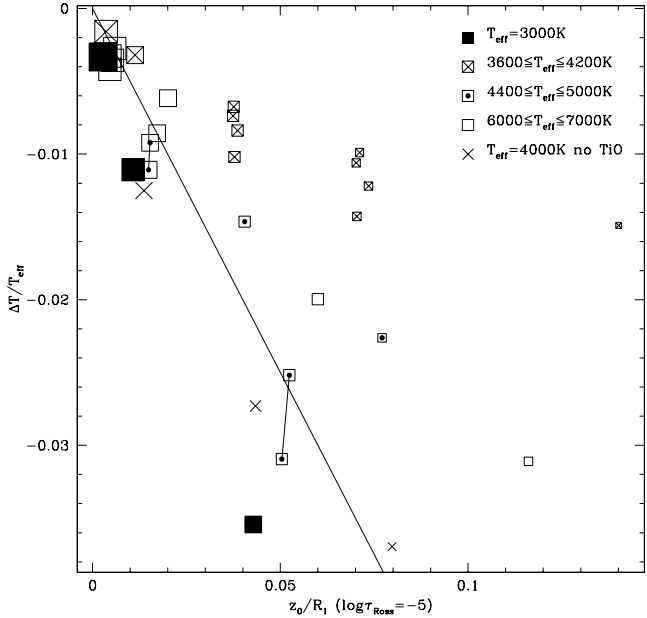}}
\\
\\
\caption{The difference in the surface temperature (at log\,$\tau_{\rm Ross}=-5$) between spherical and
plane-parallel models with
$T_{\rm eff}$ ranging from 3000\,K to 7000\,K (with bigger symbols the higher log$\,g$)
and metallicities [Me/H]=0.0, except for two models with [Me/H]$=-1.0$ which are 
attached with thin lines to corresponding [Me/H]=0.0 models (the lower symbols in the pairs representing the more metal-poor
models). Three models at $T_{\rm eff}=4000\,$K have also been calculated with TiO absorption neglected and are denoted by
crosses -- it is seen
that the positions of these models depart very significantly from the corresponding standard models. The relation
according to Eq.\,(\ref{eq403}) is shown as a straight line.
} 
\label{deltaTdiff}
\end{figure} 

The effects of sphericity on the temperature structures of the models are shown in Fig.\,\ref{deltaTdiff}
and compared with the approximate expression in Eq.\,(\ref{eq403}). The agreement is satisfactory, 
though certainly not perfect, which is 
partially explained by the approximations made in deriving it. As important are the effects of 
the wavelength dependence of $\kappa_\lambda$, which weights the mean intensity $J_\lambda$ and the 
Planck function $B_\lambda$ in the radiation equlibrium integrals of Eq.\,(\ref{eq29}) differently for 
different models, depending in particular on the molecular absorption, and resulting in the large scatter
in Fig.\,\ref{deltaTdiff}.   

\begin{figure}
\resizebox{\hsize}{!}{\includegraphics{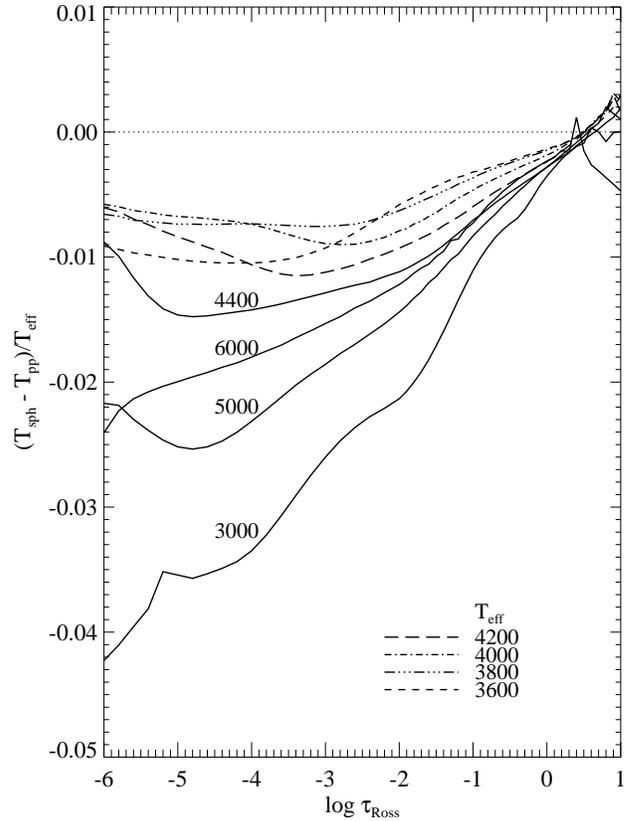}}
\\
\\
\caption{
The difference between temperature structures of spherically symmetric and corresponding plane-parallel models with 
log$\,g=1.0$, $M=1.0\,M_\odot$,
[Me/H]=0.0, and $T_{\rm eff}$ ranging from 3000\,K to 6000\,K. The complex variation of the sphericity effects in the upper layers
when TiO sets in as a heating agent at $T_{\rm eff}\approx 4000\,$K is illustrated.} 
\label{tdiff2}
\end{figure} 

\begin{figure}
\resizebox{\hsize}{!}{\includegraphics{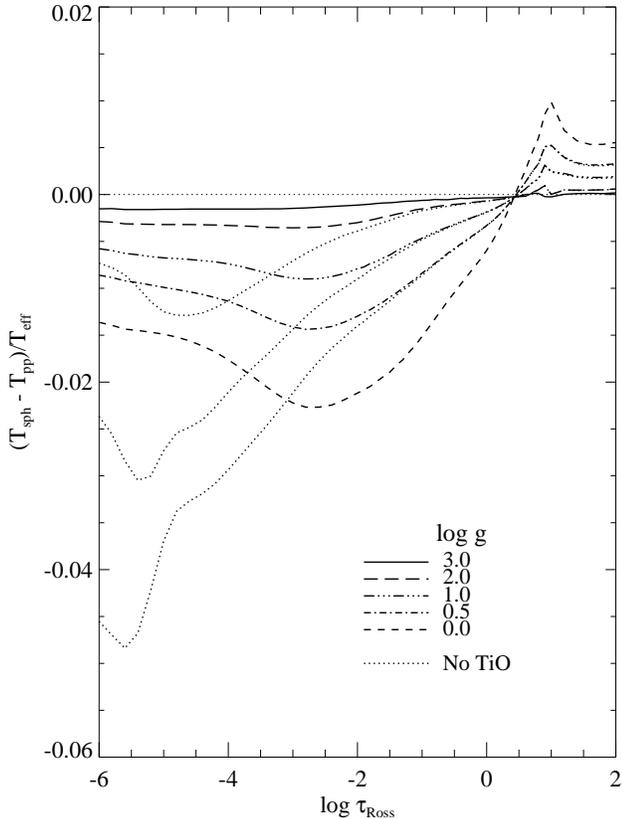}}
\caption{
The difference between temperature structures of spherically symmetric and corresponding plane-parallel models with 
$T_{\rm eff}=4000\,$K and 
different log$\,g$. $M=1.0\,M_\odot$ and 
[Me/H]=0.0, and with the TiO absorption alternatively switched on and off (dashed lines, the gravities of these can be
traced via the models to which they adhere at depth). The complex variation of the sphericity effects in the upper layers
due to TiO  is illustrated.} 
\label{tdiff}
\end{figure} 

Now turning to the detailed model results, the sphericity effects on $\Delta T/T$ are shown as a function of $\tau_{\rm Ross}$ for a sequence
of supergiant models with different effective temperatures in Fig.\,\ref{tdiff2}. Obviously, the effects shift with $T_{\rm eff}$.
First being relatively small for the hottest models, they considerably increase for $T_{\rm eff}\sim 5000\,K$, then diminish again to reach a minimum at 
about 3800\,K and finally again increase strongly as $T_{\rm eff}$ goes towards 3000\,K. This behaviour may be interpreted as a coupling
between the sphericity effects and the effects of molecular blanketing. 
The coupling may show up as a positive or negative feed-back. Sphericity cools the upper layers, relative to a plane-parallel model,
and the molecular formation increases. Around 5000\,K this leads to a strong surface cooling by CO. For cooler models TiO forms, which
conversely heats the upper atmosphere. Finally, H$_2$O takes over as a dominating opacity source, and that again cools the model surface. That 
this explains the behaviour can be tested by blocking out different molecular opacities from the model calculations. In Fig.\,\ref{tdiff} we thus
show a sequence of models at $T_{\rm eff}=4000\,{\rm K}$ with different log$\, g$ and with and without TiO. It is clear that the differential
sphericity effects above $\tau_{\rm Ross}$ are changed significantly by the TiO absorption. From this figure we also see that the sphericity effects
in the temperature structure get significantly greater than 1\% for models with log$g \le 1.0$. However, if turbulent pressures are allowed for this
value may be higher, according to Eq.(\ref{eq7}).

\begin{figure}
\resizebox{\hsize}{!}{\includegraphics{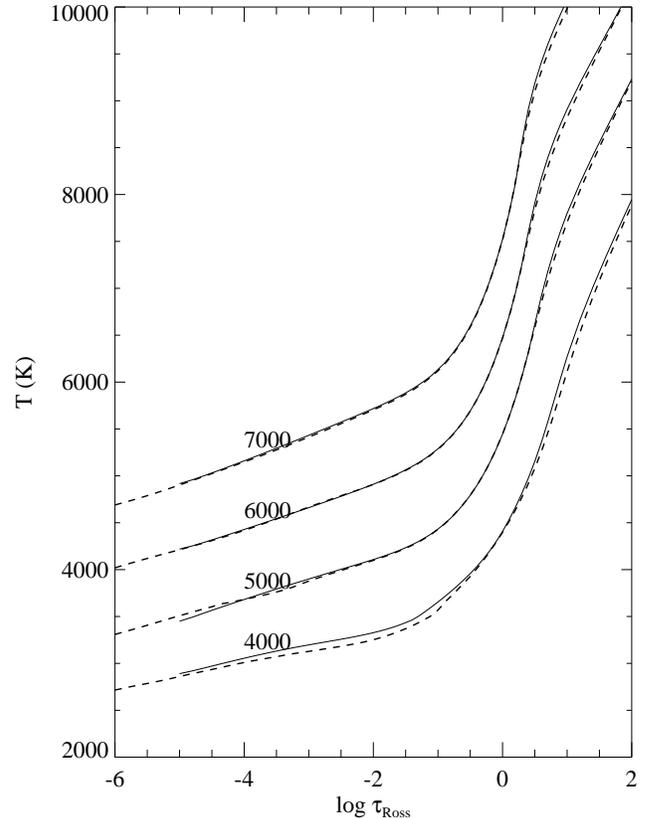}}
\caption{MARCS model atmospheres (solid) for dwarf stars with log$\,g=4.5$, [Me/H]=0.0 and corresponding models from the 
Castelli \& Kurucz ODFNEW grid (dashed). The curves are labeled with relevant values of $T_{\rm eff}$. } 
\label{krzplotdwarfs}
\end{figure}

\begin{figure}
\resizebox{\hsize}{!}{\includegraphics{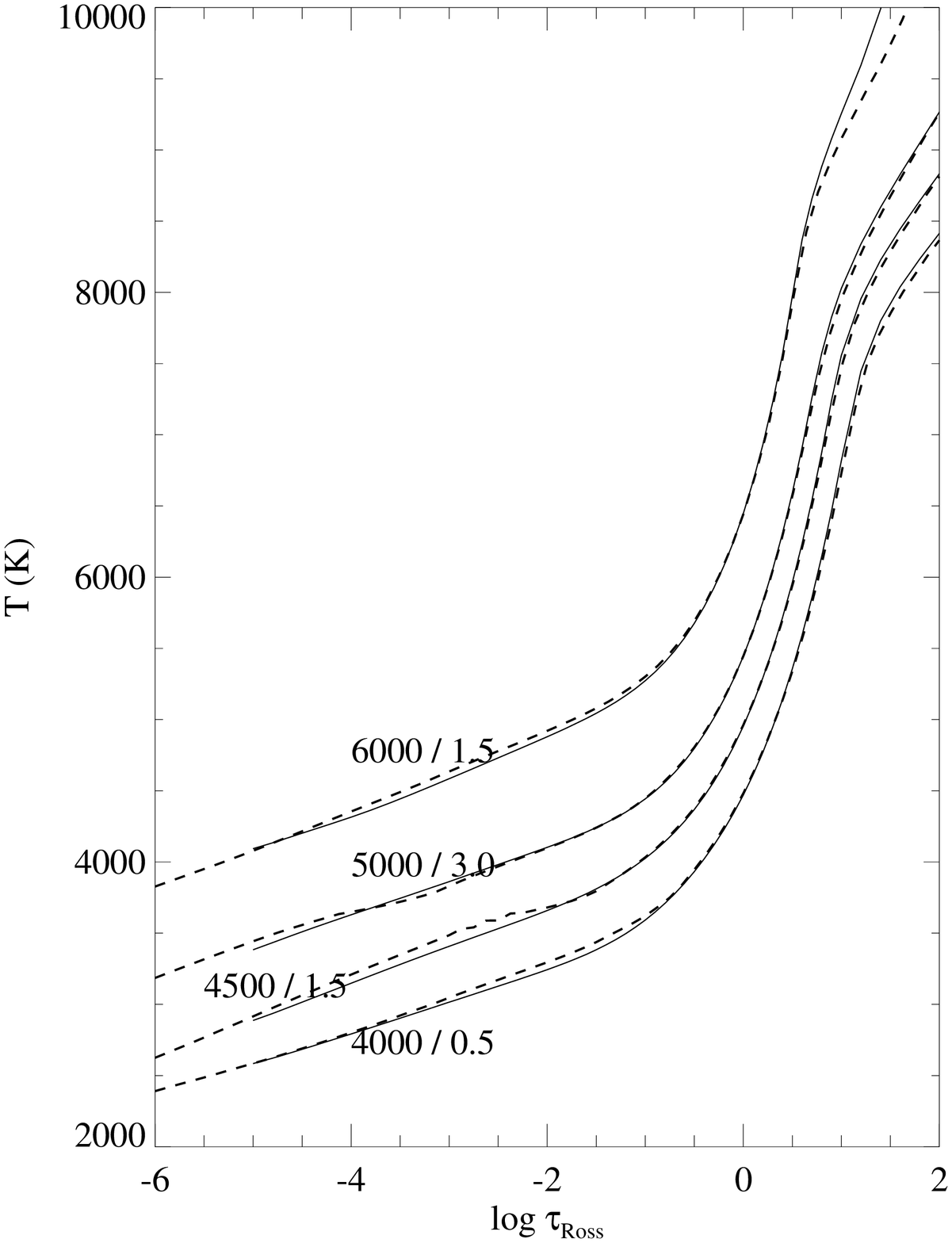}}
\caption{MARCS model atmospheres for giants and supergiants with [Me/H]= 0.0 (solid) and corresponding models from the 
recent Castelli \& Kurucz grid (dashed). The curves are labeled with relevant values of $T_{\rm eff}$ and log$\,g$.} 
\label{krzplotgiants}
\end{figure}   

\section{Comparison with other model grids}
We shall here make some comparison of MARCS grid models with the models of other existing grids, but confine this 
comparison to the temperature structures -- some additional comments on calculated fluxes will be be made in
subsequent papers of this series. 

\begin{figure}
\resizebox{\hsize}{!}{\includegraphics{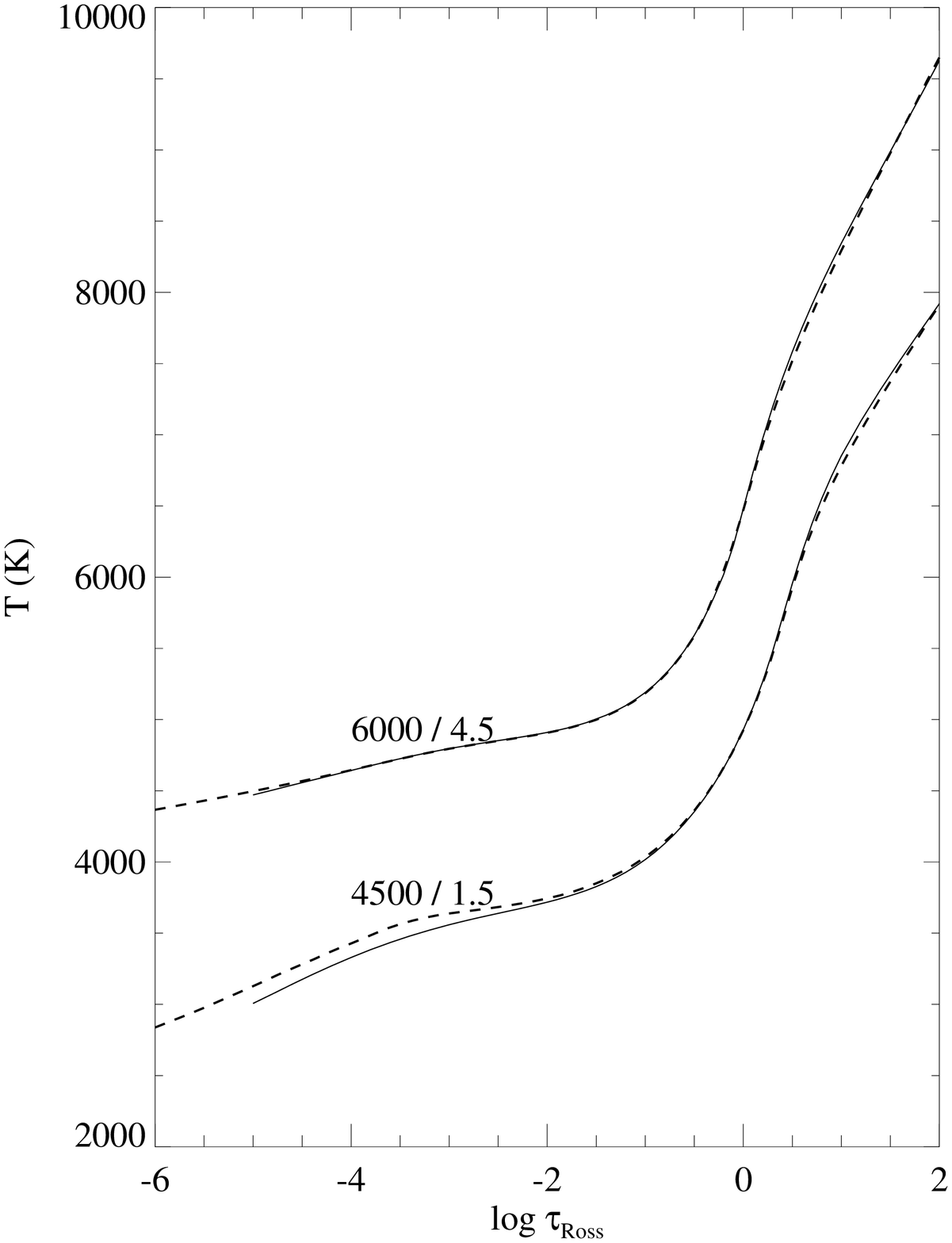}}
\caption{MARCS model atmospheres with [Me/H]=-2.0 (solid) and corresponding models from the 
{Castelli \&} Kurucz ODFNEW grid (dashed). The curves are labeled with relevant values of $T_{\rm eff}$ and log$\,g$.} 
\label{krzplotmetminus2}
\end{figure}  

The most extensive and widely used grid of models for most of our parameter space are the plane-parallel models 
of R. Kurucz and collaborators (available on {\it http://kurucz.harvard.edu}). Among the different sets of models 
published there, it is particulary relevant
to compare with those that are calculated by Castelli \& Kurucz 2003) 
with a standard mixing-length theory (without "convective overshoot"), which also seem to produce 
more consistent model parameters when applied to real data (Castelli, Gratton \& Kurucz, 1997).
Some sample comparisons of model structures are shown in Figs.\,\ref{krzplotdwarfs}, \ref{krzplotgiants} and
\ref{krzplotmetminus2}, where we have chosen Castelli \& Kurucz ODFNEW models to compare with. It is seen that the
agreement in the temperature structures is almost perfect for the models of solar-type dwarf stars of different
metallicities. This is true also for models of early M dwarfs. For the models of giants and supergiants, the
agreement is again good, although the MARCS models tend to be some 10-80 K cooler in the surface layers from 
$\tau_{\rm Ross}\le 10^{-2}$. The ageement is also very satisfactory if pressures or densities are intercompared.
In view of the fact that these two grids of models are made with two totally independent numerical methods and
computer codes, with independent choices of basic data (although Kurucz's extensive lists of atomic line transitions
are key data underlying both grids), this overall agreement is both satisfactory and gratifying.

\begin{figure}
\resizebox{\hsize}{!}{\includegraphics{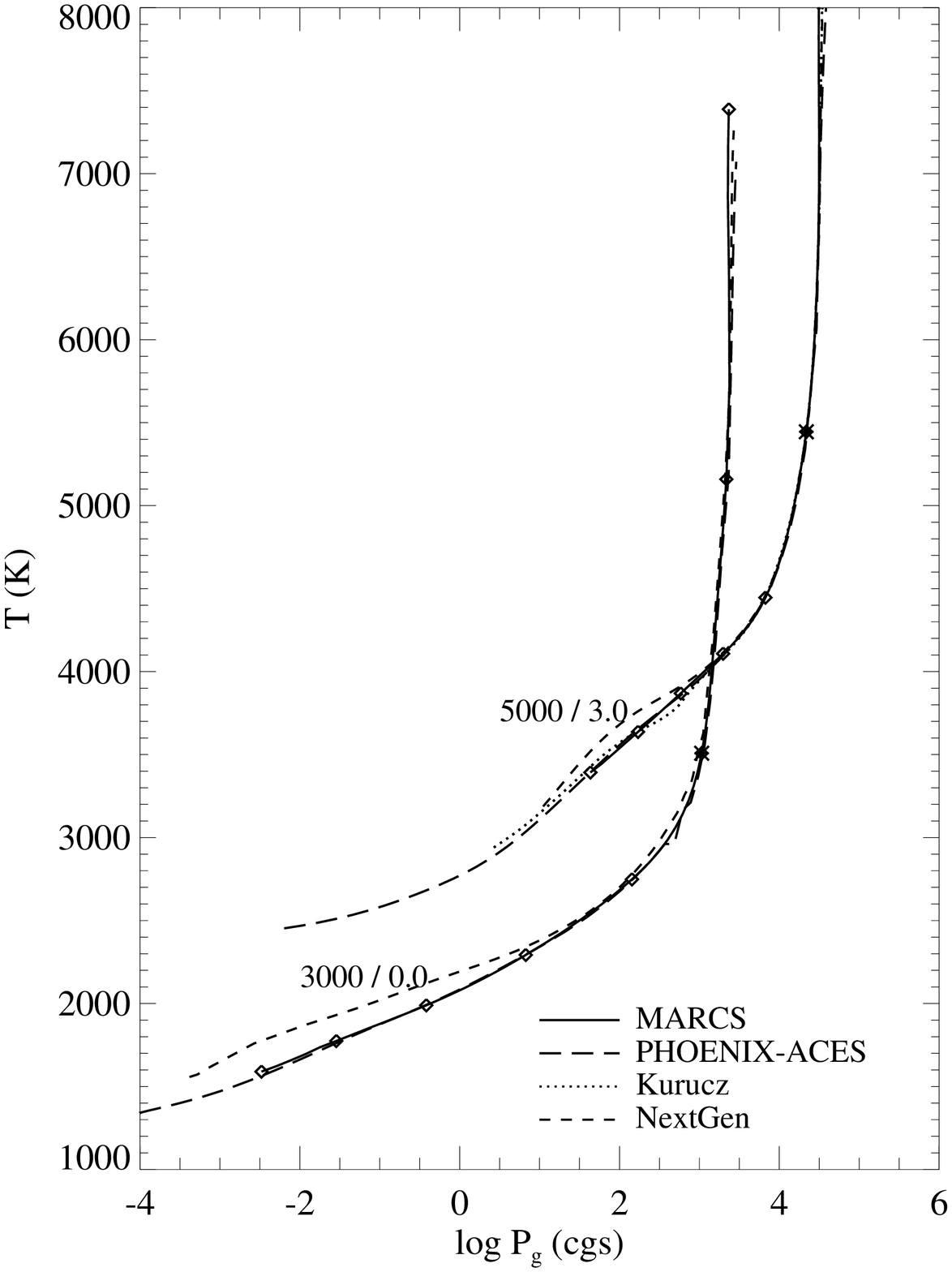}}
\label{ngplotc}
\caption{
Model atmospheres with solar abundances, from the MARCS, NextGen and PHOENIX-ACES-2008 grids and (for $T_{\rm eff}=5000\,$K) the 
ODFNEW Castelli \& Kurucz grid. log$\,g$ is 3.0 (5000\,K) and 0.0 (3000\,K), respectively. The MARCS and NextGen models are
spherically symmetric with $M=5M_\odot$ while the Kurucz model is plane-parallel. Along the MARCS curve
every decade in $\tau_{\rm Ross}$ is marked by a $\diamond$ sign, and $\tau_{\rm Ross}=0.0$ by a $\times$. 
The general impression of a better agreement between MARCS and Castelli \& Kurucz
models than between MARCS and NextGen models is also shown if the temperature  is plotted 
relative to optical depth.} 
\end{figure} 
\begin{figure}
\resizebox{\hsize}{!}{\includegraphics{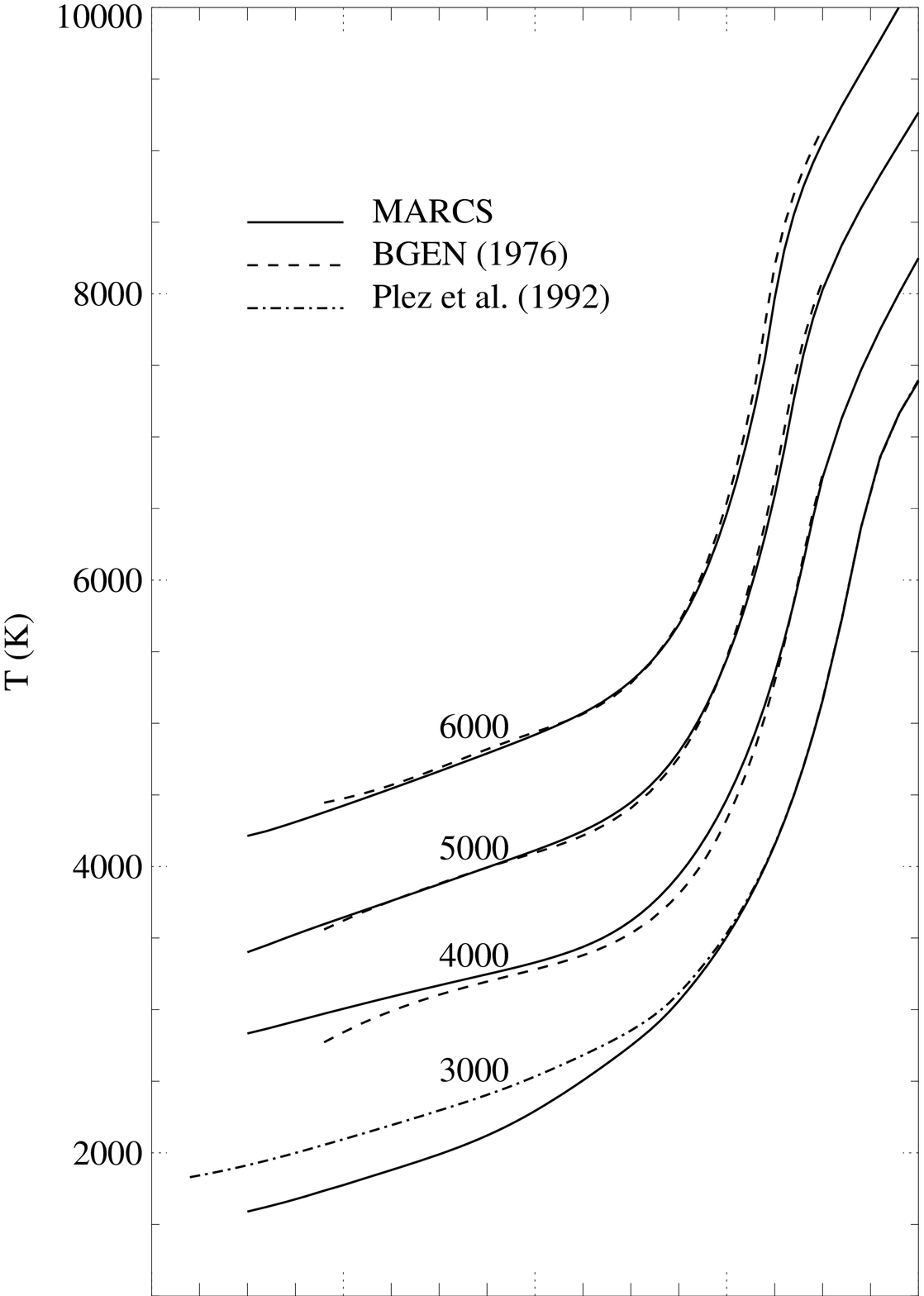}}
\vspace{5mm}
\caption{
Model atmospheres with solar abundances, from the present MARCS grid compared with corresponding MARCS models from
Bell et al. (1976, BGEN, dashed) and with one model from Plez (1992). 
log$\,g$ is 3.0 and the geometry is plane-parallel for the 6000\,K, 5000\,K and 4000\,K models; the 3000\,K models 
are spherically symmetric with $M=5$M$_\odot$ and log\,$g=0.0$.} 
\label{mxplot}
\end{figure} 

The situation was somewhat less satisfactory for models from the NextGen grid of spherically symmetric models, calculated with the
PHOENIX code (Hauschildt et al.
1999, models obtained by private communication in 1996). In the inner photospheres the agreement is good but again a difference systematically appears at the surface, 
starting around $\tau_{\rm Ross} \approx 10^{-2}$, with the MARCS models 
beeing cooler, but now by as much as typically 250\,K. (As is seen in Fig.\,\ref{ngplotc}, the agreement between the Castelli \& Kurucz and 
the MARCS models is generally much better than with the NextGen models.) However, more recent LTE models (PHOENIX-ACES-2008) obtained 
by Hauschildt with an updated 
version of the PHOENIX code (Hauschildt, private communication) and which are also based on Kurucz line data, agree very well with the MARCS models. The basic reason for this improvement in the fit seems to be due to changes in
molecular opacities. Again, the present overall fit is very satisfactory, in view of the independence of the methods of the MARCS and PHOENIX
teams. 

A number of comparisons with models made by previous versions of MARCS have also been performed. In Fig.\,\ref{mxplot} we thus compare with
models of G and K giants (Bell et al. 1976) calculated with the first version of MARCS. It is seen that the present temperature structures agree 
quite well with the older ones in spite of the
much more incomplete line absorption data of those models. (An exception from this is seen in the outer layers of the 4000\,K model, where the
TiO absorption, missing in the BGEN models, contributes a surface heating of the more recent model).
The pressures in the present models are, however, significantly lower which
reflects their more heavy opacities. The good agreement in the temperature structures must thus at least partially be 
fortuitous, e.g. reflecting that the line absorption missing in the 1976 models 
is not heavily biassed towards the ultraviolet or the red spectral regions relative to the absorption that is in.
In Fig.\,\ref{mxplot} we also show
one example where a spherically symmetric model for a late-type M supergiant is compared with a previous spherically
symmetric MARCS model of Plez et al. (1992). The considerable 
differences in the upper photosphere are due to the new and improved molecular data used now and its more detailed 
representation, in particular for the H$_2$O absorption.  

\section{Interpolation between the models}
It is useful to be able to interpolate the model structures to
a set of parameters {$T_{\rm eff}$, $\log~g$, and [Me/H]} different 
from those tabulated in the grid. Such an interpolation is not completely
straightforward, as strong non-linearities appear in the behaviour of some 
of the thermodynamic variables when stellar parameters are varied.
One program was developed by T. Masseron (2006) and extensively tested
by him on a previous grid of MARCS models encompassing the following range of parameters:
3800~K$<T_{\rm eff}<$7000~K, $0.0<{\rm log}~g<5.0$, $-4.0<{\rm [Me/H]}<0.0$.
It interpolates  $T, P_{\rm g}, P_{\rm e}, \kappa_{\rm Ross}$, and the geometrical depth
as a function of $\tau_{\rm Ross}$.
The interpolated model must lie inside a complete cube of existing models 
in the parameter space ($T_{\rm eff}$, $\log~g$, [Me/H]), and the interpolation
is optimized to account for non-linearities in the grid.
With the actual grid parameter steps, maximum errors in the interpolated quantities
should remain below 0.25\% for $T$ and a few \% for $P_{\rm g}$ and $P_{\rm e}$, whereas
$\kappa_{\rm Ross}$ and the column density ({\it rhox}) are more difficult to interpolate.
The program, with a detailed user manual, is available on the MARCS web site:
{\sl http://marcs.astro.uu.se}.

An interpolation routine, applied to structures, as well as fluxes and colours of models by
Kurucz and collaborators, has 
also been described by Nendwich et al. (2004), see also Valenti \& Fischer (2005).  

\section{Conclusions}
It will certainly take additional time before model atmospheres for late-type stars are regularly
constructed with physically more adequate assumptions than the standard ones -- of spherical symmetry, 
mixing-length convection and LTE -- adopted here. Until then, standard models will be the common
choice in e.g. abundance analyses. Here we have presented such a grid of standard
models to the state of the art. It includes model atmospheres for stars of 
spectral types from F to M, as well as carbon stars and Pop II stars of different metallicities.
The assumptions, physical data used and numerical methods have been described, and a number of general properties
of the models have been illustrated. In particular, we have discussed the effects of blanketing and sphericity 
in some detail and tested several of the existing approximate analytical estimates of these effects
and found them to work quite well. Also, already existing grids, such as the
grid of plane-parallel models by Castelli \& Kurucz and the new PHOENIX grid, have been compared with and the
agreement has been found to be excellent in the region of overlapping parameters. This is particularly
gratifying in view of the fact that the different approaches, as regards numerical methods and codes and
to a considerable degree also the selection of data, are independent. We may then basically regard the
calculation of standard 1D LTE model atmospheres as becoming routine, but important improvements in line absorption
data are still needed, and certainly not easily achieved. This is also true for data for dust formation and
dust opacities.  

Further presentations and discussions of particular properties of these 
models will appear in subsequent papers in this series. Hopefully, these will serve also as starting
points for further studies of the adequacy and shortcomings of the standard models, and as a stimulus
for endeavours to set higher standards in the art of modelling stellar atmospheres in the future. 
\begin{acknowledgement}
We thank the Swedish Science Research Council and the Centre National de la Recherche Scientifique for long-standing support during the 
years. BP thanks Sveneric Johansson for his hospitality for several years at the Department for Physics at Lund.
Previous
collaborators in the development of the MARCS code are also thanked: Roger A Bell, Martin Asplund,
John Brett, Leen Decin, Ulf Ekberg, G\"oran Hammarb\"ack, Dan Kiselman, Michelle Mizuno-Wiedner, Olle Morell, Nils Olander, Mikael Saxner
and Nicole van der Bliek, as well as contributors and critics  
of important atomic and molecular data:  Jan Alml\"of, Bernhard Aringer, Paul Barklem, Manuel Bautista, Aleksandra Borysow, Alan Irwin,
Per Jensen, Sveneric Johansson, Alain Jorissen, Bob Kurucz, David Lambert, Mats Larsson,
Bo Lindgren, Nik Piskunov, Francois and Monique Querci, Jaques Sauval, Per Siegbahn and Georg Ole S\o rensen. Peter Hauschildt is thanked 
for providing {\it PHOENIX} models for comparison and Thomas Masseron for providing his interpolation routine for MARCS models. 
Martin Asplund, Ulrike Heiter, Bob Kurucz and Jeff Linsky are thanked for valuable comments on the manuscript.  

\end{acknowledgement}

\end{document}